\documentclass[3p]{elsarticle}

\usepackage{xcolor,textcomp}
\usepackage{graphicx,hyperref,epstopdf,mathtools}
\usepackage{amsmath,amssymb,multirow,array,gensymb,bm}
\usepackage{fancyhdr,float}
\usepackage[none]{hyphenat}
\usepackage{enumitem,relsize}
\usepackage{footmisc,mathtools}
\usepackage{soul,subfigure,longtable}
\usepackage[utf8]{inputenc}
\usepackage{lineno}
\usepackage{tikz}
\usepackage{orcidlink}
\usepackage{caption}
\usepackage{nomencl}
\usepackage{comment}
\usepackage{natbib}

\makenomenclature

\newcolumntype{P}[1]{>{\centering\arraybackslash}p{#1}}
\newcolumntype{M}[1]{>{\centering\arraybackslash}m{#1}}

\biboptions{sort&compress}

\AtBeginDocument{%
  \def\thefnote{\myfnsymbol{fnote}}}
\makeatletter
\def\myfnsymbol#1{\expandafter\@myfnsymbol\csname c@#1\endcsname}
\def\@myfnsymbol#1{\ifcase #1\or $\dagger$\or $\#$\else \@ctrerr\fi}
\def\fntext[#1]#2{\g@addto@macro\@fnotes{%
   \refstepcounter{fnote}\elsLabel{#1}%
   \def\thefootnote{\thefnote}
   \global\setcounter{footnote}{\c@fnote}%
   \footnotetext{#2}}}
\makeatother

\begin{document}

\begin{frontmatter}


\title{Quantifying the effect of resonant amplitude and frequency of phononic material vibrations on the coupled fluid-structure interaction dynamics in separated aerodynamic flows}

\author[inst1]{Vinod Ramakrishnan\orcidlink{0000-0002-5588-876X}}
\author[inst2]{Arturo Machado Burgos\orcidlink{0000-0001-8327-7790}}
\affiliation[inst1]{organization={The Grainger College of Engineering, Department of Mechanical Science and Engineering},
  addressline={University of Illinois Urbana-Champaign},
  postcode={Illinois, 61801},
  city={Urbana},
  country={USA}}
  \affiliation[inst2]{organization={The Grainger College of Engineering, Department of Aerospace Engineering},
  addressline={University of Illinois Urbana-Champaign},
  postcode={Illinois, 61801},
  city={Urbana},
  country={USA}}
\author[inst1]{Sangwon Park\orcidlink{0009-0009-0129-5414}}
\author[inst2]{Andres Goza\orcidlink{0000-0002-9372-7713}}
\author[inst1]{Kathryn H. Matlack\corref{cor1}\orcidlink{0000-0001-7387-2414}}
\cortext[cor1]{Corresponding author}
\ead{kmatlack@illinois.edu}


\begin{abstract}
Phononic materials (PMs) with engineered resonances have been leveraged for fluid-structure interaction (FSI) with fluid flow instabilities, yielding beneficial outcomes such as transition delay, stabilized hypersonic boundary layers, and increased aerodynamic lift. Prior PM-FSI studies primarily identify spatio-temporal flow scales of interest and choose PM structural parameters producing structural dynamics conducive for FSI. However, a fully-coupled FSI system generally produces complex coupled dynamics that is not accurately captured by studying either physical system in isolation. In this context, our prior work established behavioral parameters that govern the coupled PM-FSI dynamics in a separated aerodynamic flow over a limited parameter range. Adopting this framework, this paper explores strongly-coupled high-fidelity PM-FSI simulations over a broader range of two behavioral parameters---truncation resonance frequency and displacement amplitude---to establish their quantitative (linear/cubic) relations to the coupled frequency, lift force, and circulation in the coupled system response. In addition, the results indicate the presence of distinct FSI regimes, depending on the proximity of the truncation resonance frequency or its sub-/super-harmonics to the vortex-shedding frequency. FSI dynamics ranging from multi-/single-frequency dynamics, downshifted coupling frequency due to fluid-added mass effects, generation of non-linear harmonics to convergence of FSI dynamics to the rigid plate case are observed. These results reiterate the importance of the PM frequency and amplitude in determining the coupled FSI dynamics, and the proposed quantitative relations provide a new pathway for designing PMs for aerodynamic flow control to achieve beneficial outcomes, e.g., lift force enhancement.
\end{abstract}


\end{frontmatter}

\nomenclature[01]{$\alpha$}{Angle of Attack}
\nomenclature[02]{$l$}{Plate length}
\nomenclature[03]{$t$}{Time}
\nomenclature[04]{$\Gamma$}{Plate surface}
\nomenclature[05]{$\Gamma_\mathrm{CS}$}{Compliant Section}
\nomenclature[06]{$\chi$}{Displacement of the midpoint of $\Gamma_\mathrm{CS}$ (interface mass of PM)}
\nomenclature[07]{$m_1,m_2$}{Lumped masses in the diatomic PMs}
\nomenclature[08]{$k,k_\mathrm{g}$}{Inter-mass spring stiffness and the grounding spring stiffness in the diatomic grounded PM}
\nomenclature[09]{$a$}{Spatial periodicity of diatomic PM}
\nomenclature[10]{$F_\mathrm{CS}$}{Force acting in the normal direction to the plate surface in $\Gamma_\mathrm{CS}$}
\nomenclature[11]{$C_{l}$}{Lift force acting in the $y$-direction on $\Gamma$}
\nomenclature[12]{$\gamma$}{Circulation in the flat-plate wake}
\nomenclature[13]{$\square^\mathrm{R}$}{$F_\mathrm{CS},C_{l},\gamma$ signals recorded in the flow over a rigid inclined flat plate configuration. Absence of superscript denotes signals recorded in FSI simulations.}
\nomenclature[14]{$\square_\mathrm{mean}$}{Mean values of $\chi,F_\mathrm{CS},C_{l},\gamma$ signals recorded at $t\geq20$}
\nomenclature[15]{$\square_\mathrm{amp}$}{Peak-to-peak oscillation amplitudes of $\chi,F_\mathrm{CS},C_{l},\gamma$ signals recorded at $t\geq20$}
\nomenclature[16]{$f$}{Frequency}
\nomenclature[17]{$f_\mathrm{VS}$}{Vortex-shedding frequency (global attractor)}
\nomenclature[18]{$f_\mathrm{TR}$}{First truncation resonance frequency}
\nomenclature[19]{$f_\mathrm{TR2}$}{Second truncation resonance frequency}
\nomenclature[20]{$f_\mathrm{PB}$}{Pass band resonance frequencies}
\nomenclature[21]{$f_0$}{Reference frequency}
\nomenclature[22]{$f_\mathrm{TR}^\mathrm{FSI}$}{Coupled frequency of the FSI system}
\nomenclature[23]{$\Delta f_{\%}$}{Frequency shift in FSI compared to prescribed $f_\mathrm{TR}$}
\nomenclature[24]{$A_\square$}{Fourier transform amplitudes}
\nomenclature[25]{$\mathrm{N}$}{Number of unit cells in the PM}
\nomenclature[26]{$n_\mathrm{d}$}{Number of masses within a unit cell of the PM}
\nomenclature[27]{$\mathbf{M},\mathbf{K}$}{Mass and Stiffness matrices}
\nomenclature[28]{$k_\mathrm{eff}$}{Static effective stiffness}
\nomenclature[29]{$\lambda$}{Displacement amplitude envelope}
\nomenclature[30]{$m_\mathrm{UC}$}{Total mass of a single PM unit cell}

\printnomenclature

\section{Introduction}\label{sec:Introduction}

Fluid-Structure Interaction (FSI) of various fluid flows with phononic materials (PMs) has emerged as a vibrant area of research in the last decade, owing to the ability to precisely engineer PM vibrational characteristics such as the frequency of vibration~\citep{KushwahaPRL1993,BastawrousJASA2022,HasanPRSA2019,HasanJEL2024,RamakrishnanJSV2025}, the absolute vibrational amplitude~\citep{RamakrishnanJSV2025}, and the direction of vibration~\citep{LeeSMS2023,MiniaciJAP2021}, that can interact with equivalent spatio-temporal and pressure amplitude scales in the flow. Specifically, PM dynamic behaviors such as evanescent wave propagation in phononic band gaps~\citep{KushwahaIJMPB1996}, truncation resonances~\citep{BastawrousJASA2022,HasanPRSA2019,HasanJEL2024}, defect resonances~\citep{JoMAMS2022,RamakrishnanJSV2025}, and behavior as an acoustic diode~\citep{SchmidtJAP2025} have proven effective for FSI. 

A straightforward approach to exploring meaningful PM-FSI is to identify dominant spatio-temporal scales of flow structures/instabilities---such as Tollmien–Schlichting waves~\citep{hussein2015flow,WilleyJFS23,MichelisPoF2023,BarnesAIAA2021,SchmidtJAP2025}, Karmann vortex streets~\citep{KeoghArXiV2025,WibergAIAA2025}, hypersonic shockwave/boundary-layer instabilities~\citep{NavarroMatter2025}, and turbulent streaks and rollers~\citep{LinAIAA2024,LinArXiV2026}---and then choose PMs with conducive resonant frequencies, amplitudes and fluid force-PM velocity phase relations to interact with these flow processes. Using this approach, recent PM-FSI studies have demonstrated beneficial outcomes such as delay of laminar-to-turbulent transition, suppression of flow vortices, stabilization of hypersonic flow and turbulent drag reduction, respectively. Though these studies highlight the budding potential of PMs for FSI and flow control applications, a central feature of the chosen problems---which allows for quick iteration and decouples the PM design process from the changing flow conditions---is that the frequency, wavelength, and topology of the targeted flow structures/instabilities remain unchanged between the coupled PM-FSI and the fluid flow over a rigid plate configurations. However, this is generally not the case in FSI, as fluid flows are nonlinear dynamical systems that adapt their flow characteristics in response to external perturbations. Said differently, the coupled flow-structure interaction system~\citep{benjamin1960effects,landahl1962stability} is nonlinear, and its dominant spatial-temporal scales will be different than those of the constituent flow and structural dynamical systems unto themselves. Therefore, casting PM parameters non-dimensionally (relative to characteristic flow parameters), and in ways that govern the holistic fluid-PM system~\citep{RamakrishnanJFS2026} is an important next step in developing PMs with designed FSI.

Canonical FSI systems~\citep{riley1988compliant,Gad-el-Hak1996compliant,williamson2004vortex,sarpkaya2004critical,shelley2011flapping} such as vortex-induced vibrations in flow past an elastic cylinder, flow past a visco-elastic membrane in a bounded/unbounded channel, have often involved complex and counterintuitive dynamics. Nevertheless, using simple parameters such as a representative structure to fluid mass ratio, or stiffness ratio, researchers have been able to successfully distinguish distinct FSI regimes, systematically tune the complex system dynamics, and identify the FSI mechanisms at play. Drawing parallels to our prior work~\citep{RamakrishnanJFS2026}, we proposed four PM behavioral parameters---the effective stiffness, $k_\mathrm{eff}$, the truncation resonance frequency, $f_\mathrm{TR}$, the amplitude envelope, $\lambda$, and the unit cell mass, $m_\mathrm{UC}$, that when varied independently produce distinct PM-FSI dynamics such as non-linear harmonic generation, shift in coupling frequency, and shift from broad band to narrow band dynamics. PM-FSI simulations across different behavioral parameter ranges produced distinct FSI behaviors past a PM-embedded inclined flat-plate with a latent vortex-shedding instability. From these results, we infer that the $\{f_\mathrm{TR},\lambda\}$ are the two most important PM control parameters that determine the nature of interaction and synchronization between the PM and flow dynamics. Note that the force-displacement phase has also emerged as an important parameter dictating the PM-FSI in recent studies~\citep{MichelisPoF2023,WilleyJFS23,BrotnowSSRN2026}. However, given the absence of a strong (not latent) flow instability, the flow dynamics lock-on to one of the dominant PM resonances, rendering the force-displacement phase $\approx-\pi/2$ in all FSI simulations explored in \citet{RamakrishnanJFS2026}. Therefore, the force-displacement phase is not used as control parameter in the earlier~\citep{RamakrishnanJFS2026} or current PM-FSI study.

Our prior work~\citep{RamakrishnanJFS2026} examined only a limited range of PM behavioral parameters and provided a qualitative analysis of the emergent FSI behavior. However, an exhaustive sweep of the key behavioral parameters, identification of distinct FSI regimes, and establishment of quantitative relations between these independent parameters and the coupled behavior will provide a robust pathway to precisely engineer the PM-FSI dynamics. Therefore, using the PM behavioral parameter framework proposed in \citet{RamakrishnanJFS2026}, this article characterizes distinct coupled PM-FSI regimes and provides detailed scaling relationships between the dominant behavioral parameters and key output quantities such as coefficient of lift, coupled frequency, and circulation. We begin with a similar analysis to \citet{RamakrishnanJFS2026} (the Fourier spectral analysis of the FSI simulations in Fig.~\ref{fig:FIG2}) but for a much broader range of frequencies, and subsequently, explore quantitative relations between the dominant PM behavioral parameters and the coupled output quantities, providing a predictive capability for precisely determining the vortex-shedding characteristics in an aerodynamic PM-FSI setting. For PM-FSI simulations, we choose a constant $\{k_\mathrm{eff},m_\mathrm{UC}\}$ and comprehensively vary the $\{f_\mathrm{TR},\lambda\}$ to characterize the different vortex shedding regimes. The PM truncation resonance frequency is varied in the sub-harmonic ($f_\mathrm{TR}\approx0.5f_\mathrm{VS}$) to the super-harmonic ($f_\mathrm{TR}\approx2f_\mathrm{VS}$) range of the latent vortex-shedding frequency, $f_\mathrm{VS}$, inherent to the inclined rigid flat-plate configuration. The amplitude envelope, $\lambda$ is varied in the $(0,\lambda_\mathrm{max})$ range, where the lower limit defines a perfectly rigid subsurface, and the upper limit corresponds to the amplitude envelope of a single mass-spring element with an equivalent $\{k_\mathrm{eff},f_\mathrm{TR}\}$, that localizes all the vibrational energy at the fluid-PM interface. Multiple high-fidelity PM-FSI simulations choosing different $\{f_\mathrm{TR},\lambda\}$ values are performed and their impact on important flow quantities are assessed.

The remainder of the article is organized as follows. Sec.~\ref{sec:Overview} briefly summarizes the FSI problem configuration, and provides an overview of the scope of FSI simulations. Sec.~\ref{sec:Results} discusses key insights from various fully coupled FSI simulations, with PMs parameterized using the behavioral parameters, $\{f_\mathrm{TR},\lambda\}$. Sec.~\ref{sec:EmpRel} explores quantitative relations between the important flow parameters and the PM behavioral parameters. Finally, Sec.~\ref{sec:Conclusion} presents a summary and the key takeaways of this article.

\section{Overview of the PM-FSI configuration}\label{sec:Overview}

\begin{figure*}[t!]
    \centering
    \includegraphics[width=\textwidth]{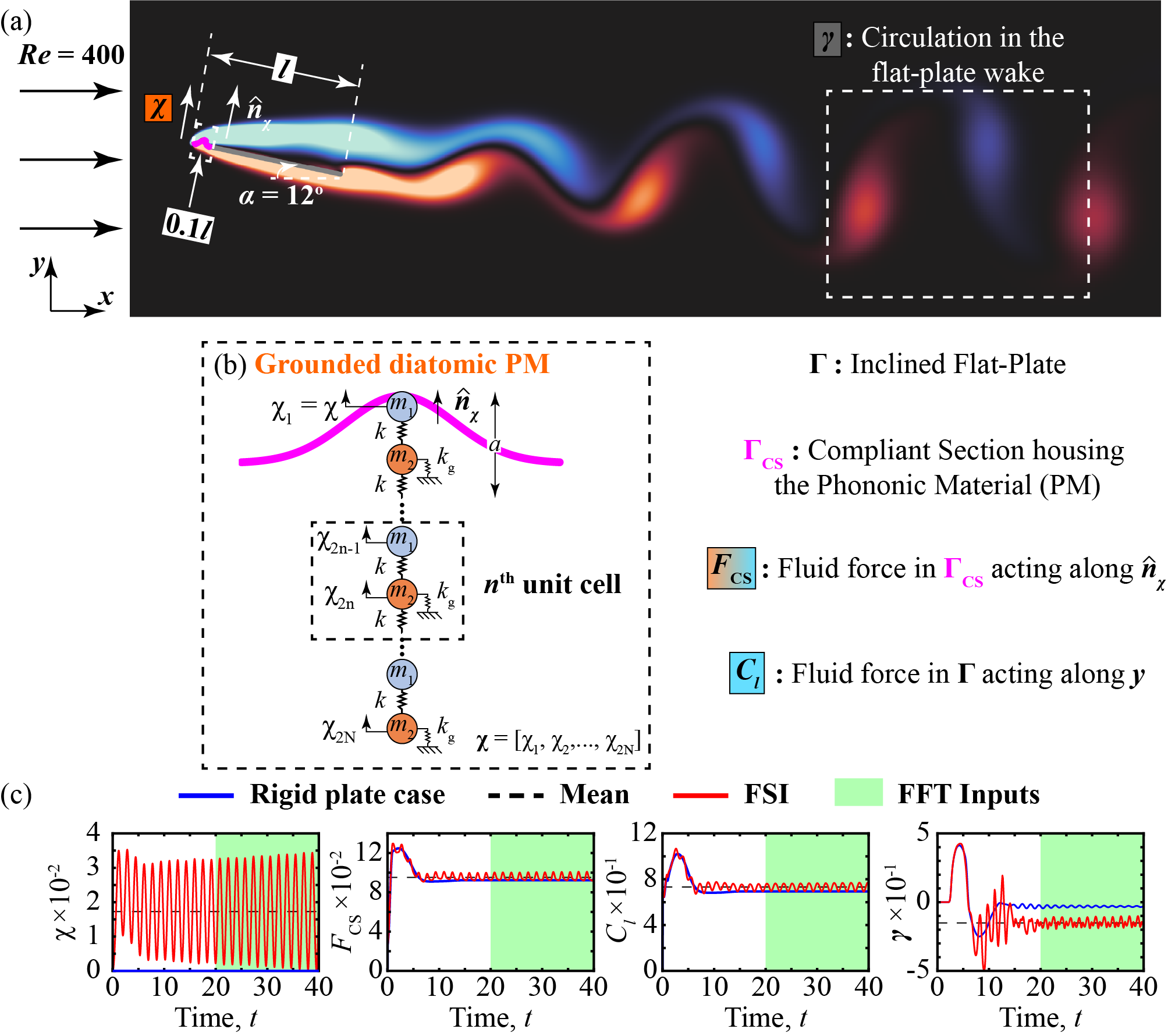}
    \caption{Fluid-PM FSI Configuration. (a) Flow at $Re=400$ over an inclined flat-plate embedded with a PM subsurface and the FSI-induced vortex-shedding behavior observed in the flat-plate wake. (b) Lumped mass-spring representation of the grounded diatomic PM. (c) Temporal signals obtained from strongly coupled FSI simulations (left to right): PM displacement, $\chi$; Force in the compliant section, $F_\mathrm{CS}$; Lift force, $C_{l}$; Circulation in the flat-plate wake, $\gamma$. }
    \label{fig:FIG1}
\end{figure*}

We adopt the phononic material (PM) integrated inclined-flat plate configuration described in \citet{RamakrishnanJFS2026}, which we summarize here for self-containment of presentation. The current study utilizes the same configuration and identified behavioral parameters, but studies a more extensive range of the behavioral parameters to identify behavioral regimes and quantify scaling relationships between behavioral parameters and output quantities such as frequency and lift of the coupled system. 

\subsection{FSI of a PM-embedded flat-plate with an aerodynamic flow}\label{sec:FSI_Configuration}

Fig.~\ref{fig:FIG1}a shows a schematic of our FSI setup that consists of an incompressible flow ($Re=400$) past a two-dimensional, infinitesimally thin flat plate of chord length $l$ inclined at $\alpha=12^\circ$ to the oncoming freestream (from left to right). For flow over a completely rigid inclined flat-plate, these parameters represent flow conditions where the flow separates slightly at the trailing edge; that is, the flow is still steady but near the onset of an unsteady instability. A slight increase in angle of attack, for example, would trigger indefinite periodic vortex shedding. This motivates these problem parameters in the FSI simulations considered in this study.

The numerical simulations are performed using the high-fidelity strongly coupled immersed boundary algorithm of \citet{goza2017strongly}, which has been validated on multiple FSI problems. The fluid occupies a rectangular domain (light gray shaded region in Fig.~\ref{fig:FIG1}a),$\{x,y\} \subset \mathbb{R}^2$, the plate surface is denoted by $\Gamma$, and a compliant section (CS), $\Gamma_\mathrm{CS} \subset \Gamma$, is located near the leading edge extending 0.1$l$ in length along plate surface. This compliant portion is illustrated in purple. The remainder of the flat-plate, $\Gamma \setminus \Gamma_\mathrm{CS}$ is kinematically rigid (dark gray section of the plate). To effectively capture the strong interaction between the PM-aerodynamic flow and avoid complex coupling between the PM mass-spring model and the compliant section (purple plate section), the CS dynamics are taken to be fully determined by the PM dynamics. Consequently, the midpoint displacement of the compliant section, $\chi(t)$, is equated to the displacement response of the PM interface mass, $\chi_1(t)$. Using the CS midpoint as the apex point, a spatial Gaussian function is defined that kinematically prescribes the deformation of the entire CS ($\Gamma_\mathrm{CS}$). A detailed overview of the Gaussian parametrization of the compliant section is available in \citet{RamakrishnanJFS2026}.

We adopt a lumped mass-spring description of a discrete grounded diatomic PM~\citep{RamakrishnanJFS2026} ($n_\mathrm{d}=2$ degrees of freedom per unit-cell) to model the PM dynamics. This PM description features a one-dimensional (1D) chain of two masses, $\{m_1,m_2\}$, elastic springs of stiffness, $k$, that connect adjacent masses, and grounding elastic springs of stiffness, $k_\mathrm{g}$, that connect each alternate mass, $m_2$ to a fixed ground (Fig.~\ref{fig:FIG1}b). We consider the momentum equations for the flow, the continuity equation enforcing conservation of mass for incompressible flow, the matrix PM equation, and the no-slip boundary condition ensuring that the velocity of the flow and the plate match at every material point in both the rigid and compliant section of the plate~\citep{RamakrishnanJFS2026}. These equations are fully coupled using the immersed boundary formulation; that is, the nonlinear flow behavior is directly related to the internal PM dynamics, and consequently to the CS motion. This coupled collection of equations is
\begin{gather}
\frac{\partial \bm{\mathsf{u}}}{\partial t}+\bm{\mathsf{u}}\cdot\nabla \bm{\mathsf{u}}=-\nabla \mathsf{p} +\frac{1}{Re}{\nabla}^2\bm{\mathsf{u}}+\int_{\Gamma}\bm{\mathsf{f}}(\tilde{s},t)\delta(\bm{\mathsf{X}}(s(\tilde s,t),t)-\bm{{x}}) d\tilde{s},
\label{eqn:NS} \\
\nabla\cdot\bm{\mathsf{u}}=\mathsf{0}, \label{eqn:continuity} \\
\mathbf{M}\,\frac{d^2\bm{\chi}}{dt^2} + \mathbf{K}\,\bm{\chi}(t) = \bm{\mathbf{f}}_{\text{PM}}(t), \label{eqn:mass_spring} \\
\int_{\Omega}\bm{\mathsf{u}}(\bm{{x}}, t)\delta(\bm{{x}}-\bm{\mathsf{X}}(s(\tilde s,t),t))d\bm{{x}}=\frac{d\,\bm{\mathsf{X}}(s(\tilde s,t),t)}{dt} \quad \forall \; \bm{\mathsf{X}}\in \Gamma_\mathrm{CS}. \label{eqn:no-slip}
\end{gather}
where, $\bm{\mathsf{u}}$ and $\mathsf{p}$ are the flow velocity and pressure fields, $\bm{\mathsf{X}}(s,t)$ is the coordinate position of the plate (incorporating the spatially Gaussian and temporally evolving motion of the compliant section ($\Gamma_\mathrm{CS}$), $\{\mathbf{M},\mathbf{K}\}$ represent the mass and stiffness matrices, $\bm{\chi}=[\chi_1,\chi_2,\dots,\chi_{2\mathrm{N}}]^\mathrm{T}$ represents the displacement vector collating the displacements of all the $2\mathrm{N}$ PM masses ($\mathrm{N}$ unit-cells), and, $\mathbf{f}_\mathbf{PM}=[F_\mathrm{CS},0,\dots,0]^\mathrm{T}$ represents the force vector collating all external forces applied to the PM masses. See reference \citep{RamakrishnanJFS2026} for full details on how the flow loading, compliant section motion, and internal PM dynamics are coupled.

The governing  equations are first recast from their primitive variable form into a streamfunction-vorticity formulation. This transformation eliminates the pressure variable and ensures that the continuity equation (Eq. \ref{eqn:continuity}) is satisfied everywhere, drastically improving computational efficiency. The new formulation is then discretized in space, yielding a semi-discrete system of time-continuous equations. To advance the system in time, the spatially discretized momentum equation (Eq. \ref{eqn:NS}) is integrated using an Adams-Bashforth scheme for the nonlinear terms, a Crank-Nicolson method for the diffusive term, and an implicit treatment of the surface stress term. This ensures that the no-slip condition (Eq. \ref{eqn:no-slip}) is enforced to within machine precision at each time step. Concurrently, the PM dynamics (Eq. \ref{eqn:mass_spring}) are advanced using a second-order implicit Newmark scheme. Newton's method is then employed to solve the resulting nonlinear algebraic system of equations arising from the strongly coupled FSI between the flow and the PM. Finally, a block-LU decomposition is applied to the linearized system of equations for an efficient iterative solution process \citep{goza2017strongly}. To avoid redundancy, we only provide a brief summary of the immersed boundary formulation for the PM-aerodynamic flow FSI configuration. A detailed overview of the full numerical setup, and of grid convergence tests that demonstrate suitability of the method, are in \citet{RamakrishnanJFS2026}.

\subsection{PM behavioral parameters and the latent vortex-shedding behavior}\label{sec:PMBehavioral_&_VS}

The rigid inclined flat-pate at an angle of attack, $\alpha=12^\circ$ shows a latent vortex-shedding behavior concentrated at a frequency, $f_\mathrm{VS}$. As mentioned in Sec.~\ref{sec:FSI_Configuration}, the system is dynamically near a bifurcation leading to indefinite periodic vortex shedding. At the current angle of attack, the global attractor is a time-steady state. For example, providing an impulse to the flow state at some time would trigger transient vortex shedding that would decay exponentially with an associated frequency of $f_{\mathrm{VS}}=0.6256$. (At slightly higher angle of attack or Reynolds number, the impulsive force would produce an exponential growth in vortex-shedding intensity until nonlinearities saturate the behavior to a persistent limit cycle, with frequency near that of $f_\mathrm{VS}$.) Because of this intrinsic but latent timescale in the flow system, $f_{\mathrm{VS}}$ is used as the representative flow timescale against which PM-FSI dynamics are evaluated in this study.  

Our prior work~\citep{RamakrishnanJFS2026} proposed four behavioral FSI parameters. For completeness in this article, we briefly summarize and give intuition for these parameters. Mathematically, these parameters were proposed as:
\begin{enumerate}
    \item The effective stiffness, $k_\mathrm{eff}$, governs the mean static displacement at the fluid-PM interface where it is subject to a near-constant lift force in the given aerodynamic flow setting.
    \item The unit cell mass, $m_\mathrm{UC}$, tunes the proximity of $f_\mathrm{TR}$ to other PM resonances. In the dimensionless form introduced in~\citet{RamakrishnanJFS2026}, it also presents a representative structural mass relative to a characteristic added mass the flow provides.
    \item The truncation resonance frequency, $f_\mathrm{TR}$, is the natural frequency of the first resonance mode of vibration of the PM designed in this study to be within the ``band gap'' of the PM's dispersion curve~\citep{DeymierSpringer2013}. This key structural timescale can be systematically tuned relative to the characteristic flow frequency $f_\mathrm{VS}$.
    \item The displacement envelope~\citep{RamakrishnanJSV2025,RamakrishnanJFS2026}, $\lambda$, quantifies the time rate of increase of the displacement amplitude of the interface PM mass, when excited at a given structural resonance frequency. This key structural amplitude scale can be systematically tuned to vary the strength of coupling between the PM and the flow.
\end{enumerate}

\subsection{Scope of FSI Simulations and Analysis}\label{sec:FSI_Scope}

Our prior work~\citep{RamakrishnanJFS2026} identified $f_\mathrm{TR}$ and $\lambda$ as the key parameters in determining the coupling frequency and the fluid-PM coupling strength, over a small range of these parameters. Motivated by these outcomes, the main objective of this study is to make use of the identified behavioral parameters~\citep{RamakrishnanJFS2026}, and perform an extensive numerical experiment campaign across a broad range of the most dynamically important behavioral parameters: the truncation resonance frequency and the displacement amplitude envelope. 

The behavioral parameters for FSI simulations are chosen as follows. Thirteen distinct truncation frequencies, $f_\mathrm{r}=f_\mathrm{TR}/f_\mathrm{VS}=0.419,0.466,0.5,0.513$, $0.6,0.7,0.839,0.932,1,1.025,1.2,1.4,2$; in the range of the sub-harmonic to the super-harmonic of the $f_\mathrm{VS}$ are chosen. The range of the amplitude envelope varies with $f_\mathrm{TR}$ ($\lambda \propto f_\mathrm{TR}$). Therefore, given an arbitrary $\lambda_0$ close to $\lambda_\mathrm{max}$, we choose $\{\lambda_{1},\cdots,\lambda_{6}\}=2\lambda_0/7,3\lambda_0/7$, $4\lambda_0/7,5\lambda_0/7,6\lambda_0/7,\lambda_0$, uniformly distributed across the range, $(0,\lambda_\mathrm{max})$. Note that though the absolute value of $\lambda_0$ (and consequently $\lambda_{1,2,\cdots,6}$) changes for different $f_\mathrm{TR}$ (see Tab.~\ref{tab:FSI_PM_Sims}), the choice of uniformly distributed $\lambda\in(0,\lambda_\mathrm{max})$ allows us to qualitatively compare the FSI results for different $f_\mathrm{TR}$. The effective stiffness value is chosen as $k_\mathrm{eff}=5.4533$ to target a static mean displacement of $\chi_\mathrm{mean,ref}=0.017$ (same as \citet{RamakrishnanJFS2026}), and focus on the dynamic fluid-PM interaction across all 78 FSI simulations (13 distinct $f_\mathrm{TR}$ values $\times$ 6 distinct $\lambda$ values). Finally, the range of the unit cell mass also varies with $f_\mathrm{TR}$ ($m_\mathrm{UC} \propto 1/f_\mathrm{TR}^2$). So, the unit cell mass is chosen as $m_\mathrm{UC}=1.8/f_\mathrm{r}^2$, for the different $f_\mathrm{TR}$ values. Note that $m_\mathrm{UC}$ is maintained constant for all 6 simulations with different $\lambda$ values at the given $f_\mathrm{TR}$.

Further, we also conduct 13 additional FSI simulations that integrate the single mass-spring element in the compliant section, that represent the maximum amplitude envelope, $\lambda_\mathrm{max}$, for a given $f_\mathrm{TR}$. The single mass-spring configuration is mechanically equivalent to a PM with a highly localized truncation mode shape (i.e., $\kappa_I\rightarrow\infty$). Therefore, it represents the upper limit of the range of PM amplitude envelope, $\lambda_\mathrm{max}$~\citep{RamakrishnanJFS2026}. This follows from the fact that though the amplitude envelope and truncation mode shape are independent quantities, they are implicitly correlated with a high $\lambda$ reflecting a more localized mode shape (high $\kappa_I$) and vice versa.

Therefore, a total of 91 strongly-coupled FSI simulations (78 PM-FSI + 13 single mass-spring-FSI) are performed and their results analyzed. See Tab.~\ref{tab:FSI_PM_Sims} for the comprehensive list of behavioral parameters used in the FSI simulations.

\section{Results of FSI Simulations}\label{sec:Results}
A detailed analysis of the numerical FSI simulation results is presented in this section. We quantify several variables in the PM, flow, and their interface, and the representative time domain results are given in Fig.~\ref{fig:FIG1}c (coupled response for FSI cases with $f_\mathrm{r}=1$ and high $\lambda=\lambda_6$). Specifically, we analyze the PM displacement $\chi$, the interface fluid force, $F_\mathrm{CS}$, the total lift force, $C_{l}$, acting in the $y$-direction (in-plane normal direction to the freestream flow) along the entire $\Gamma$, and the near-wake flow circulation, $\gamma$, calculated within the the region $\{x,y\}:[2.5,5.5]\times[-0.5,0.5]$, indicated in dark gray on Fig.~\ref{fig:FIG1}c. Fig.~\ref{fig:FIG1}c also plots the representative $\{\chi,F_\mathrm{CS},C_{l},\gamma\}$ signals obtained in the rigid plate case. As evident from these plots, the FSI between the PM and the aerodynamic flow significantly alters these quantities compared to the rigid plate case.

\begin{figure*}[t!]
    \centering
    \includegraphics[width=\textwidth]{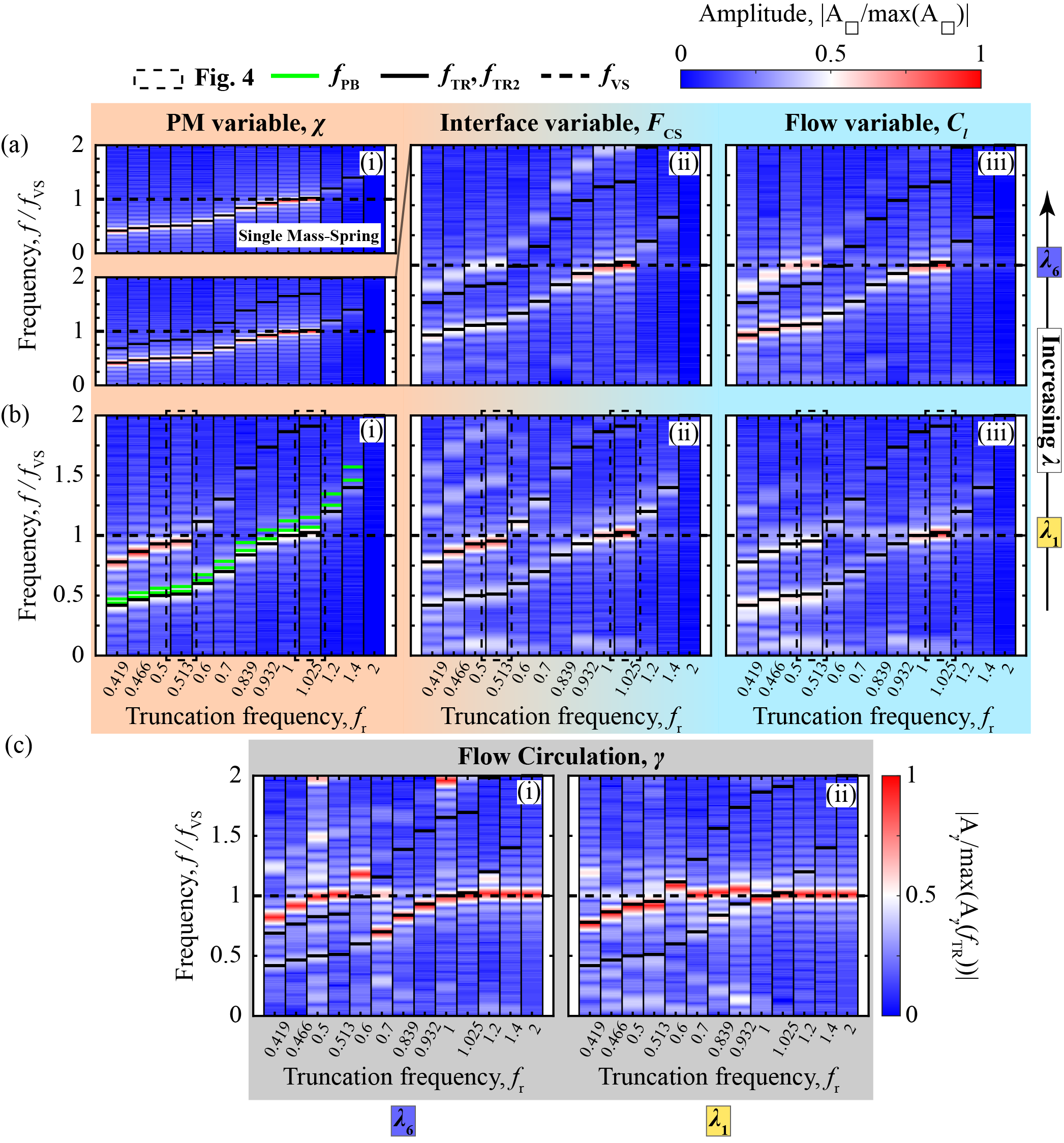}
    \caption{Fourier frequency spectra in the post-transient state ($t\geq20$). (a-b) (i) Surface displacement, $\chi$ (PM variable), (ii) interface fluid force, $F_\mathrm{CS}$ (interface variable), and (iii) lift force, $C_{l}$ (flow variable) signals from all select PM and single mass-spring FSI configurations. FSI configurations featuring PMs with (a) $\lambda=\lambda_6$, and (b) $\lambda=\lambda_1$. (c) The flow circulation, $\gamma$, observed in the shaded wake region depicted in Fig.~\ref{fig:FIG1}a, for PMs with (i) $\lambda=\lambda_6$, and (ii) $\lambda=\lambda_1$. Structural parameters are given in Table~\ref{tab:FSI_PM_Sims}. See supplementary files MOV.1 and MOV.2 for a visualization of the spatio-temporal evolution of flow vorticity. (Note that we calculate the Fourier spectra after subtracting the mean value for $t\geq20$ from each signal, to eliminate low frequencies that can skew the spectral amplitude distributions. $\mathcal{F}(\square):\square=\square(t)-\square_\mathrm{mean}$, where $\square=\{\chi,F_\mathrm{CS},C_{l},\gamma\}$)}
    \label{fig:FIG2}
\end{figure*}

Figs.~\ref{fig:FIG2}a,b summarize the Fourier frequency spectrum of the: (i) PM displacement, $\chi$ (PM variable), (ii) force in the CS, $F_\mathrm{CS}$ (interface variable), and (iii) the lift force, $C_{l}$ (flow variable) for select FSI simulations (all 13 different $f_\mathrm{TR}$ values and 2 amplitude envelopes, $\lambda_{1,6}$). These variables are specifically chosen to connect dynamics of the PM, the fluid-PM interface and the aerodynamic flow around the flat-plate. In addition, Fig.~\ref{fig:FIG2}c summarizes the Fourier frequency spectrum of the wake circulation, $\mathbf{\gamma}$, to provide an insight of the global flow dynamics away from the fluid-PM interface.

The frequency spectra of the PM-flow variables in the FSI simulations featuring PMs with the highest ($\lambda_6$) and lowest ($\lambda_1$) amplitude envelope are represented in Fig.~\ref{fig:FIG2}a and Fig.~\ref{fig:FIG2}b. Results show that changing $\lambda$ creates qualitative differences in the $\{F_\mathrm{CS},C_{l}\}$ signals such as narrow or broadband PM responses, stronger or weaker nonlinear harmonic generation effects, for all $f_\mathrm{TR}$. 

\subsection{Frequency spectrum analysis for high $\lambda$}\label{sec:High_Lambda}

We first analyze the frequency spectrum for the cases with high $\lambda$ ($=\lambda_6$, see Figs.~\ref{fig:FIG2}a.i-iii,c.i, and MOV.1 for a visualization of the spatio-temporal evolution of the flow vorticity). The coupled results in Figs.~\ref{fig:FIG2}a, highlight that the PM-FSI dynamics concentrate spectrally close to the prescribed $f_\mathrm{TR}$ for high $\lambda$, irrespective of whether that frequency is close to the latent vortex-shedding frequency. The coupled dynamics also converge to those of a single mass-spring FSI as the $\lambda$ increases. To demonstrate this latter point, Fig.~\ref{fig:FIG2}a.i includes a split figure plotting the spectral map of the FSI dynamics when the PM is replaced by a single mass-spring with a stiffness $k=k_{\mathrm{eff}}$ and a mass $m$ chosen so that the natural frequency $\sqrt{k/m} = 2\pi f_{\mathrm{TR}}$. The displacement frequency spectra for the $\lambda_{6}$ cases, and the single mass-spring cases are visually indistinguishable. In addition, there is a strong presence of a nonlinear second harmonic, $2f_\mathrm{TR}$, in the flow that indicates a strong fluid-PM coupling. This is a consequence of the PM interacting at a high amplitude (high $\lambda$) and in a narrow band manner with the flow. Note that the Fourier spectra in each sub figure of Figs.~\ref{fig:FIG2}a.i-iii is normalized by the maximum spectral amplitude observed across all FSI simulations with $f_\mathrm{r}\in[0.419,2]$ and $\lambda=\lambda_6$. This enables a comparative analysis of the effect of (mis-)aligning the $f_\mathrm{TR}$ with the $f_\mathrm{VS}$, given a proportional level of fluid-PM coupling ($\lambda$).

Fig.~\ref{fig:FIG2}a.i shows a strong concentration of PM dynamics at the engineered $f_\mathrm{TR}$ for $f_\mathrm{r}\lesssim1$ (i.e., $0.419\leq f_\mathrm{r}\leq 1.025$), and significantly weak PM dynamics for $f_\mathrm{r}>1$ (i.e., $f_\mathrm{r}=1.2,1.4,2$). It is worth re-iterating that despite the uneven spectral distribution of flow energy across different frequencies (as seen in the forcing signal, $F_\mathrm{CS}^\mathrm{R}$ in the rigid plate scenario~\citep{RamakrishnanJFS2026}), the PM spectral response is centered around the prescribed $f_\mathrm{TR}$ in all the FSI scenarios ($f_\mathrm{r}\lesssim1$). Among these spectra, the PM responses with $f_\mathrm{r}=0.932,1,1.025$, where the resonance frequencies lie in the proximity of $f_\mathrm{VS}$, intuitively, see the highest concentration of vibrational energy.

For $0.419\leq f_\mathrm{r}\leq 0.513$ (i.e., $f_\mathrm{r}\approx0.5$) spectral peaks at intermediate amplitudes ($\mathrm{A}_{F_\mathrm{CS}}\approx0.5$, white colored peaks) are observed at both the primary and secondary harmonics of the first truncation resonance, $\{f_\mathrm{TR},2f_\mathrm{TR}\}$, in the $F_\mathrm{CS}$ signals (a proxy for the local flow near the compliant section, $\Gamma_\mathrm{CS}$). In the $C_{l}$ signal (a proxy for the flow over the entire plate, $\Gamma$) the spectral energy redistributes between the primary and secondary harmonics. Consequently, the spectral peaks in the the $C_{l}$ signals at the second harmonics for PMs with $f_\mathrm{r}=0.5,0.513$, and at the primary harmonics for PMs with $f_\mathrm{r}=0.419,0.466$, show comparable energy levels to the absolute maximum observed for PMs with $f_\mathrm{r}=1.025$. For $f_\mathrm{r}=0.5,0.513$, owing to the proximity of the second harmonics to $f_\mathrm{VS}$, the spectral energy redistributes from the primary harmonic in the local flow to the second harmonic in the flow over the plate. Conversely, a reverse energy redistribution from the second harmonic in the $F_\mathrm{CS}$ to the primary harmonic in the $C_{l}$ is observed for $f_\mathrm{r}=0.419,0.466$.

For $0.6\leq f_\mathrm{r}\leq 1.025$ a dominant primary harmonic with a relatively weak second harmonic of the truncation resonance is visible in the interface variable. As the second harmonics are far away from $f_\mathrm{VS}$, they vanish from the flow as it evolves over the plate ($C_{l}$ signal). Among all the Fourier spectra, FSI cases with $f_\mathrm{r}=0.932,1,1.025$, where the resonance frequencies lie in the proximity of $f_\mathrm{VS}$, intuitively, see the highest concentration of spectral energy in both the $\{F_\mathrm{CS},C_{l}\}$ signals, with the maximum energy concentrations observed at $f_\mathrm{r}=1.025,1,0.932$, in that order.

For $f_\mathrm{r}=1.2,1.4,2$, we observe weak spectral amplitudes in the $\{F_\mathrm{CS},C_{l}\}$ signals (Fig.~\ref{fig:FIG2}a.ii-iii). 
For these cases, the behavior of $F_\mathrm{CS}$, $C_{l}$ is comparable to that of a rigid plate (cf. the discussion in \citet{RamakrishnanJFS2026}, not described in detail here).
This indicates a lack of significant FSI coupled dynamics that causes the flow to return to the latent vortex-shedding state. This observation will be further corroborated when we quantitatively analyze the FSI simulations in Sec.~\ref{sec:EmpRel}.

Finally, the Fourier spectral analysis of the circulation, $\gamma$, (Fig.~\ref{fig:FIG2}c.i) in the plate wake, shows that the flow locks-on to primary or second harmonic frequency of $f_\mathrm{TR}$ in different FSI cases. Note that, the color contours in Figs.~\ref{fig:FIG2}c.i-ii are normalized so that the maximum energy at each frequency has a value of one. This choice is distinct from Figs.~\ref{fig:FIG2}a,b, where the contour levels are normalized by the maximum across all frequencies. This choice was made to identify a clear range of $f_\mathrm{TR}$ ($\lesssim1$) that accommodates significant PM dynamics capable of altering the flow behavior, and then subsequently study the impact of PM-FSI on the near-wake flow (away from the plate). For $0.7\leq f_\mathrm{r}\leq 1.025$ the dominant spectral peaks close to the prescribed $f_\mathrm{TR}$ indicate a flow lock-on to the primary harmonic. For $0.419\leq f_\mathrm{r}\leq 0.6$ the flow synchronizes with the second harmonic as it is closer to $f_\mathrm{VS}$. The peaks close to $f_\mathrm{VS}$ observed for $f_\mathrm{r}=1.2,1.4,2$, reflect wake unsteadiness at the $f_\mathrm{VS}$ as the PM-flow coupled dynamics are very small at these truncation frequencies, causing the flow to essentially resemble the steady rigid case (with very small unsteady perturbations).
Therefore, the results in Fig.~\ref{fig:FIG2}c.i corroborate the existence of a flow lock-on phenomena that complements our earlier observations, and a global attractor effect that influences the flow to prefer frequencies closer to $f_\mathrm{VS}$. The strong flow lock-on to $f_\mathrm{TR}$ or $2f_\mathrm{TR}$ also indicates the prominent role of the PM dynamics in determining the coupled FSI dynamics.

\subsection{Frequency spectrum analysis for low $\lambda$}\label{sec:Low_Lambda}

We now analyze the frequency spectrum for the cases with a low $\lambda$($=\lambda_1$, see Figs.~\ref{fig:FIG2}b.i-iii,c.ii, and MOV.2 for a visualization of the spatio-temporal evolution of the flow vorticity). Fig.~\ref{fig:FIG2}b.i, highlights a broadband PM response with multiple spectral peaks at the primary truncation resonance, the pass band resonances, and even the second truncation resonance for some cases. In addition to these primary resonance frequencies, weak spectral peaks at the second harmonics of the multiple PM resonance frequencies are also visible in the interface and flow variables.

For $0.419\leq f_\mathrm{r}\leq 0.513$ (i.e., $f_\mathrm{r}\approx0.5$) the second truncation resonance, owing to its proximity to $f_\mathrm{VS}$ dominates the PM dynamics compared to the primary truncation resonance. Note that the second truncation modes in the high $\lambda$ cases also lie close to $f_\mathrm{VS}$ (Fig.~\ref{fig:FIG2}a.i); however, these resonances have virtually no contribution to the PM dynamics. Therefore, proximity to $f_\mathrm{VS}$ is only a necessary but, not a sufficient condition for significant participation of the second truncation resonance in the PM dynamics. The presence of other conducive factors in the the low $\lambda$ cases contribute to the strong spectral presence of $f_\mathrm{TR2}$, and a detailed analysis investigating this phenomena is included later in Sec.~\ref{sec:fTR2}. The dominance of the second truncation resonance is also visible in the interface variable ($F_\mathrm{CS}$). In fact, the maximum spectral peaks at $f_\mathrm{TR2}$ in the $f_\mathrm{r}=0.466,0.5,0.513$ cases are comparable in magnitude to the absolute maximum observed for the $f_\mathrm{r}=1.025$ case. In addition, since the second harmonics of the primary truncation resonance ($2f_\mathrm{TR}$) are close in value to $f_\mathrm{TR2}$ (see Figs.~\ref{fig:FIG4}a.ii-iii), relatively broad spectral peaks are observed around these frequencies. In the flow variable, the primary and secondary harmonic signatures larger than $f_\mathrm{VS}$ disappear and spectral peaks relatively equal  in magnitude ($A_{C_l}\approx0.5$) are observed around the primary and secondary truncation resonances.

For $0.6\leq f_\mathrm{r}\leq 1.025$ a dominant presence of the primary truncation resonance is observed in the PM dynamics (Fig.~\ref{fig:FIG2}b.i). Unlike previous cases, second harmonic generation effect is relatively non-existent in these scenarios. In addition, for $f_\mathrm{r}=0.6,0.7$, a weak participation of the second truncation resonance is visible in the interface variable. However, in the flow variable, the primary truncation resonance dominates the flow dynamics for all $f_\mathrm{r}$. In both the interface and flow variables, relatively narrow band spectral peaks are observed close to the prescribed $f_\mathrm{TR}$, with the absolute maximum at $f_\mathrm{r}=1.025$, closely followed by the $f_\mathrm{r}=1$ case. For $f_\mathrm{r}=1.2,1.4,2$, we observe insignificant PM/flow dynamics (Fig.~\ref{fig:FIG2}b.i-iii).

Finally, the circulation (Fig.~\ref{fig:FIG2}c.ii) reveals a flow lock-on to the second truncation resonance for $0.419\leq f_\mathrm{r}\leq 0.513$, complementing the previous observations for the PM and the flow in Fig.~\ref{fig:FIG2}b. The circulation spectra for $f_\mathrm{r}\geq 0.6$ reflect the driving effect of the underlying global attractor of the flow state in the rigid setting. Therefore, we infer that the coupled FSI dynamics changes from being flow-driven to becoming more strongly coupled as $\lambda$ increases.

\subsection{Participation of the second truncation resonance in the coupled dynamics}\label{sec:fTR2}

\begin{figure*}[t!]
    \centering
    \includegraphics[width=\textwidth]{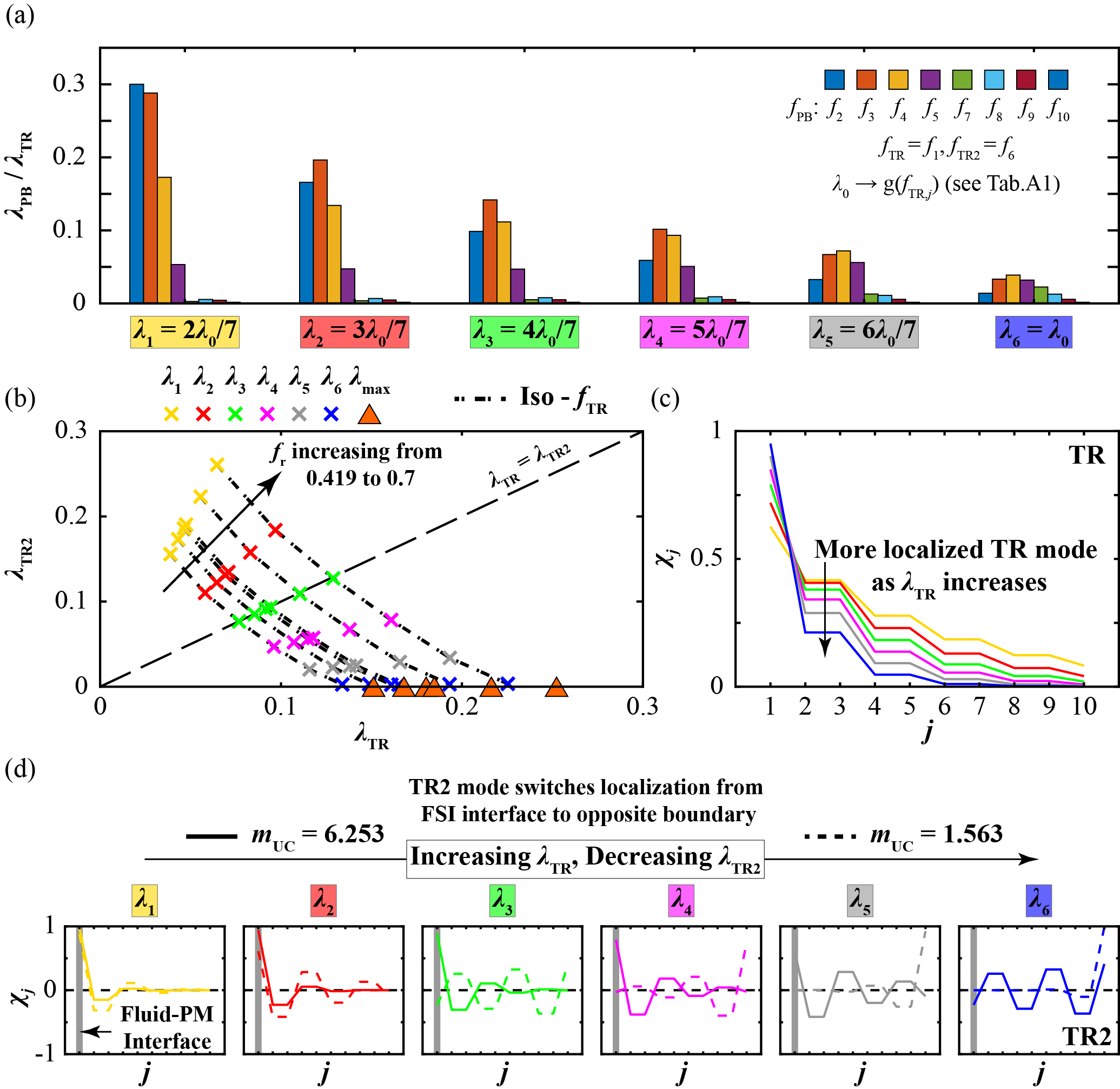}
    \caption{Relationship between amplitude envelopes associated with the primary truncation resonance ($\lambda_\mathrm{TR}$) and other PM resonances ($\lambda_\mathrm{TR2,PB}$). (a) $\lambda_\mathrm{TR}$ and $\lambda_\mathrm{PB}$ bar plots for $f_\mathrm{r}=0.5$, (b) $\lambda_\mathrm{TR}$ and $\lambda_\mathrm{TR2}$ plots for $0.419 \leq f_\mathrm{r} \leq 0.7$. (c) Evolution of the truncation resonance mode as $\lambda_\mathrm{TR}$ increases for PMs with $\{k_\mathrm{eff},m_\mathrm{UC},f_\mathrm{TR}\}=\{5.4533,6.253,0.5f_\mathrm{VS}\}$. (d) Evolution of the second truncation resonance mode as $\lambda_\mathrm{TR}$ increases for PMs with $\{k_\mathrm{eff},m_\mathrm{UC},f_\mathrm{TR}\}=\{5.4533,6.253,0.5f_\mathrm{VS}\}$ and $\{k_\mathrm{eff},m_\mathrm{UC},f_\mathrm{TR}\}=\{5.4533,1.563,0.5f_\mathrm{VS}\}$.}
    \label{fig:FIG3}
\end{figure*}

A key result in Sec.~\ref{sec:Low_Lambda} was the dominance of the second truncation resonance responded over the first truncation resonance for $0.419\leq f_\mathrm{r}\leq 0.7$. Here, we delve deeper into the structural dynamics of these corresponding PMs and the equivalent single mass-spring with the same (single) resonance frequency, $f_\mathrm{TR}$ to understand the PM-FSI results.

From \citet{RamakrishnanJFS2026}, we recall the relationship between the amplitude envelope $\lambda_\mathrm{J}$ (calculated at the fluid-PM interface mass location) and the PM resonance frequency, $f_\mathrm{J}$, associated with the J$^\mathrm{th}$ eigenmode:
\begin{equation}
\lambda_\mathrm{J}=\frac{1}{2(2\pi)^{2n_\mathrm{d}\mathrm{N}-1} f_\mathrm{J}}\cdot\left\{\frac{\mathrm{adj}\left[\mathbf{K}-(2\pi f_\mathrm{J})^2\mathbf{M}\right]}{\left[\prod_{j\neq \mathrm{J}}(f^2-f_j^2)\right]\mathrm{det(\mathbf{M}})}\right\}_\mathrm{(J,J)},
\label{eq:PM_lambda}    
\end{equation}
Here, $\{\mathbf{M},\mathbf{K}\}$ represent the mass and stiffness matrices of the PM, respectively. $n_\mathrm{d}=2$ and $\mathrm{N}=5$ denote the number of degrees of freedom within a single PM unit-cell, and the total number of unit cells in the finite grounded diatomic PM used in the FSI simulations, respectively. Eq.~\eqref{eq:PM_lambda} shows that $\lambda_\mathrm{J}$ is specific to each eigenmode, and it depends explicitly on the resonance frequency. This means that the primary and second truncation resonance of a PM will have different amplitude envelopes at their respective frequencies.

Now, given the analytical mapping~\citep{RamakrishnanJFS2026} from the behavioral parameters to the structural parameters of the PM: $\{k_\mathrm{eff},m_\mathrm{UC},f_\mathrm{TR},\lambda\}$ $\rightarrow$ $\{m_1,m_2,k,k_\mathrm{g}\}$, using Eq.~\eqref{eq:PM_lambda}, we can calculate the amplitude envelopes associated with different structural resonance frequencies, e.g., $\lambda_\mathrm{TR}$ for the primary truncation resonance, $f_\mathrm{J}=f_\mathrm{TR}$ (denoted as $\lambda$ in Secs.~\ref{sec:Introduction},\ref{sec:Overview},\ref{sec:High_Lambda},\ref{sec:Low_Lambda}), $\lambda_\mathrm{TR2}$ for the second truncation resonance, $f_\mathrm{J}=f_\mathrm{TR2}$, and $\lambda_\mathrm{PB}$ for the pass band frequencies, $f_\mathrm{J}=f_\mathrm{PB}$. 

Fig.~\ref{fig:FIG3}a plots the $\lambda_\mathrm{PB}$ of the 8 different pass band modes of the PM for the uniformly distributed $\lambda_\mathrm{TR}=\{\lambda_{1},\cdots,\lambda_{6}\}=2\lambda_0/7,3\lambda_0/7,4\lambda_0/7,5\lambda_0/7,6\lambda_0/7,\lambda_0\in(0,\lambda_\mathrm{max})$, explored in our FSI simulations. The second and third PM (pass band) resonances, $f_{2,3}$ manifest the highest $\lambda_\mathrm{PB}$ across the different $\lambda_\mathrm{TR}$. However, the maximum pass band envelopes always remain below $30\%$ of the $\lambda_\mathrm{TR}$, indicating the strong influence of the truncation mode shape on the $\lambda_\mathrm{TR}$ value by localizing the mechanical energy at the fluid-PM interface, compared to a pass band mode that disperses the energy across the PM. Moreover, the $\lambda_\mathrm{PB}$ values monotonically decrease for all the pass band modes as the truncation resonance grows stronger (i.e., as $\lambda_\mathrm{TR}$ increases). 

Fig.~\ref{fig:FIG3}b shows that the envelope associated with the second truncation resonance, $\lambda_\mathrm{TR2}$, has an inverse relationship with the envelope associated with the first truncation resonance, $\lambda_\mathrm{TR}$, i.e., low $\lambda_\mathrm{TR}$ corresponds to a high $\lambda_\mathrm{TR2}$ (yellow $\mathbf{X}$ markers), and vice versa. The single mass-spring envelopes, $\lambda_\mathrm{max}$ are also plotted in Fig.~\ref{fig:FIG3}b, to mark the upper limits ($\mathbf{\Delta}$ markers) of $\lambda_\mathrm{TR}$ for a given $f_\mathrm{TR}$. The primary reason for this inverse relationship is that the PM degree of freedom on which the $f_\mathrm{TR2}$ eigenmode localizes, changes as a function of $\lambda_\mathrm{TR}$. Figs.~\ref{fig:FIG3}c,d plot the primary and second truncation eigenmodes for $\{k_\mathrm{eff},m_\mathrm{UC},f_\mathrm{TR},\lambda\}=\{5.4533,6.253,0.5f_\mathrm{VS},[0.046,0.161]\}$, which are the parameter sets associated with Fig.~\ref{fig:FIG2}. For low $\lambda_\mathrm{TR}$, both the truncation eigenmodes are localized at the fluid-PM interface mass; however, as $\lambda_\mathrm{TR}$ increases, the $f_\mathrm{TR2}$ eigenmode starts to localize at the opposite boundary, while the $f_\mathrm{TR}$ eigenmode remains localized at the FSI interface. This shift in localization is more prominent for PMs with lower $m_\mathrm{UC}$ values. To demonstrate this, we also plot the $f_\mathrm{TR2}$ mode shapes for a PM with identical $\{k_\mathrm{eff},f_\mathrm{TR},\lambda\}$ as above, but with a smaller $m_\mathrm{UC}=1.563$. Note that the inverse relation between $\lambda_\mathrm{TR}$ and $\lambda_\mathrm{TR2}$ still holds for the PMs with lower $m_\mathrm{UC}$, with the only difference being that $\lambda_\mathrm{TR2}\approx0$ for $\lambda_\mathrm{TR}\in[0.115,0.161]$, owing to the complete transition of the $f_\mathrm{TR2}$ eigenmode localization to the opposite boundary. Therefore, for high $\lambda_\mathrm{TR}$ values, the fluid-PM interface manifests as low displacement point in the $f_\mathrm{TR2}$ eigenmode, leading to a weaker participation of the second truncation resonance in the PM dynamics (see Fig.~\ref{fig:FIG2}a.i). Alternatively, for low $\lambda_\mathrm{TR}$ values, the fluid-PM interface appears as an anti-nodal (highest displacement) point in the $f_\mathrm{TR2}$ eigenmode leading to a significant contribution in the PM dynamics (see Fig.~\ref{fig:FIG2}b.i). Therefore, high values of $\lambda_\mathrm{TR2}$ (and low $\lambda_\mathrm{TR}$) along with the higher excitation amplitudes of frequencies in the proximity of $f_\mathrm{VS}$ (as seen in the $F_\mathrm{CS}^\mathrm{R}$ signal), emerge as important factors contributing to the dominant spectral presence of $f_\mathrm{TR2}$ over $f_\mathrm{TR}$, for $0.419\leq f_\mathrm{r}\leq 0.513$ (Fig.~\ref{fig:FIG2}b).

\subsection{Shift in FSI dynamics as $\lambda_\mathrm{TR}$ changes from low to high values}

\begin{figure*}[t!]
    \centering
    \includegraphics[width=\textwidth]{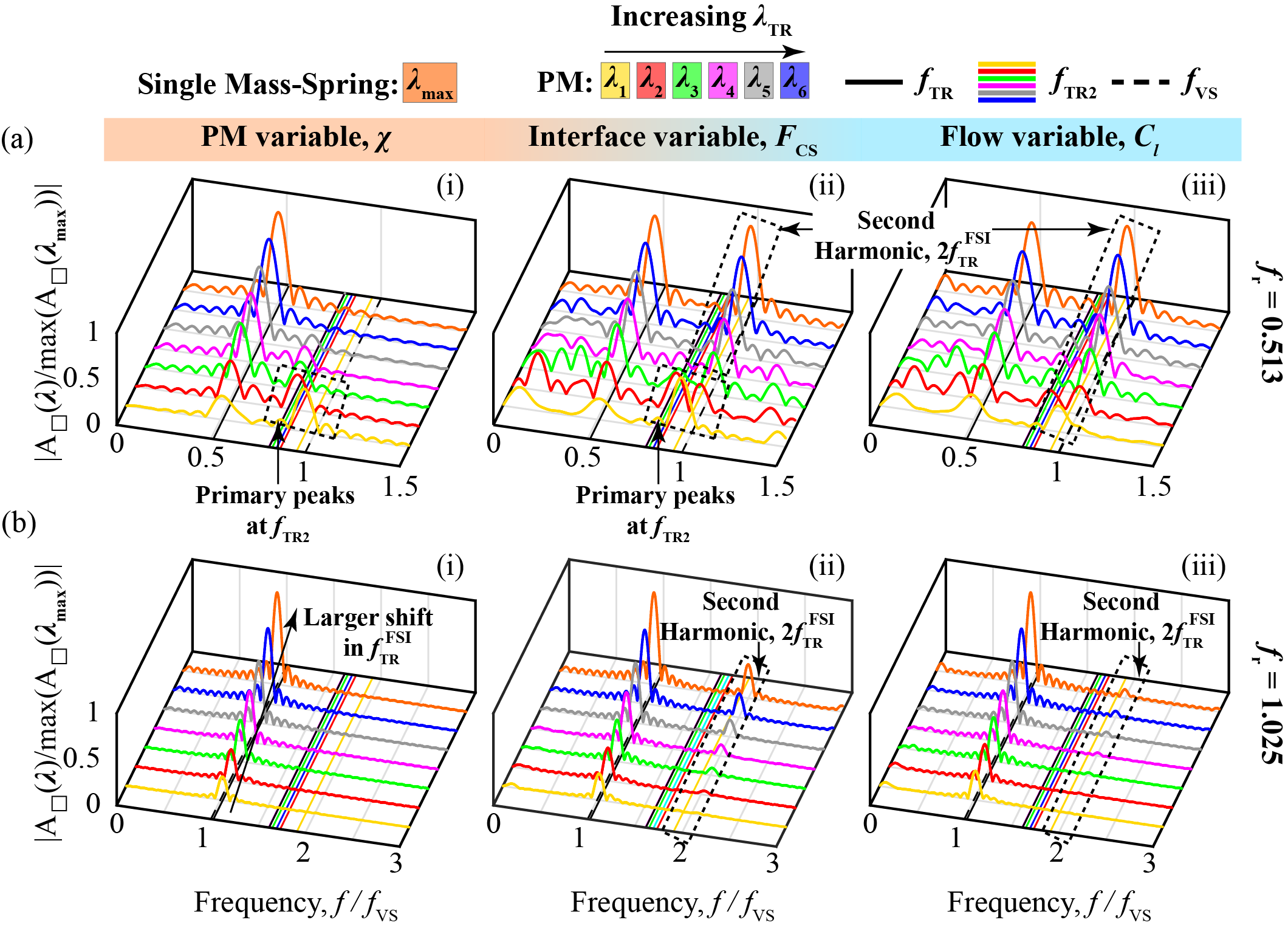}
    \caption{Vortex-shedding process as a function of different amplitude envelopes. (i) The Fourier frequency spectrum of the post-transient portion ($t\geq20$) of the surface displacement, $\chi$ (PM variable), (ii) interface fluid force, $F_\mathrm{CS}$ (interface variable), and (iii) lift force, $C_{l}$ (flow variable) signals obtained from FSI simulations highlighted in Fig.~\ref{fig:FIG2}, at (a) $f_\mathrm{r}=0.513$, and (b) $f_\mathrm{r}=1.025$.}
    \label{fig:FIG4}
\end{figure*}

Figs.~\ref{fig:FIG2}a,b revealed qualitative differences in frequency spectra for the FSI cases with the strongest ($\lambda_6$) and weakest fluid-PM coupling ($\lambda_1$), e.g., broad/narrow band responses, dominant primary/second truncation resonances, strong/weak nonlinear harmonic generation. The goal in this section is to perform a comparative analysis of the FSI results as $\lambda_\mathrm{TR}$ gradually changes from $\lambda_1$ to $\lambda_\mathrm{max}$, and explore more granular questions about the FSI dynamics: At what $\lambda_\mathrm{TR}$ value does the FSI dynamics shift from being $f_\mathrm{TR2}$ centric to being $f_\mathrm{TR}$ centric for $f_\mathrm{r}\approx0.5$? What is the effect of changing $\lambda_\mathrm{TR}$ on the coupled frequency for $f_\mathrm{r}\approx1$? In this regard, Fig.~\ref{fig:FIG4} cumulatively plots the seven FSI simulation results: 6 PM-FSI cases ($\lambda_{1}$ to $\lambda_{6}$) and one single mass-spring-FSI case ($\lambda_\mathrm{max}$), for specific truncation resonance frequencies, $f_\mathrm{r}=0.513,1.025$. These truncation frequencies are deliberately chosen as they are slightly mis-aligned with the sub-harmonic and the primary harmonic of the latent vortex-shedding frequency. The spectra in each sub figure in Fig.~\ref{fig:FIG4} are normalized by the maximum spectral amplitude among the respective simulations (i.e., the spectral amplitude in the respective single mass-spring-FSI cases). 

For $f_\mathrm{r}=0.513$, the spectral peaks in the PM displacement (Fig.~\ref{fig:FIG4}a.i) monotonically increase (decrease) close to $f_\mathrm{TR}$ ($f_\mathrm{TR2}$) as the $\lambda_\mathrm{TR}$ increases ($\lambda_\mathrm{TR2}$ decreases). The single mass-spring spectra emerges with the highest relative spectral energy concentration close to the prescribed $f_\mathrm{TR}$, as expected. Furthermore, $f_\mathrm{TR2}$ emerges as the dominant PM vibrational mode for $\lambda_\mathrm{TR}=\lambda_1 (<\lambda_\mathrm{TR2})$. The spectral peaks at $f_\mathrm{TR,TR2}$ approximately equalize for $\lambda_\mathrm{TR}=\lambda_2$ and the primary truncation resonance becomes the dominant contributor to the PM dynamics for $\lambda_\mathrm{TR}>\lambda_2$. This trend is consistent with the inverse correlation between $\lambda_\mathrm{TR}$ and $\lambda_\mathrm{TR2}$, and conforms to the fact that close to $\lambda_\mathrm{TR}=\lambda_\mathrm{3}$, the amplitude envelopes for both truncation resonances equalize (i.e., $\lambda_\mathrm{TR}=\lambda_\mathrm{TR2}$, see Fig.~\ref{fig:FIG3}b). We note that this transition point---when $f_\mathrm{TR}$ versus $f_\mathrm{TR2}$ dominates---does not exactly align with when $\lambda_\mathrm{TR2}$ becomes stronger than $\lambda_\mathrm{TR}$. This slight discrepancy could be due to unequal spectral excitation amplitudes at $f_\mathrm{TR,TR2}$ in input fluid force, $F_\mathrm{CS}$. 

The transition of PM dynamics from $f_\mathrm{TR2}$ to $f_\mathrm{TR}$ happens uniformly at $\lambda_\mathrm{TR}=\lambda_3$ for all $0.419\leq f_\mathrm{r} \leq 0.7$. All the iso-$f_\mathrm{TR}$ curves in Fig.~\ref{fig:FIG3}b look visibly similar with only the absolute $\lambda_\mathrm{TR}$ values scaling linearly with $ f_\mathrm{TR}$. Therefore, the $\chi$-spectral maps (Fig.~\ref{fig:FIG4}a.i) for all $0.419\leq f_\mathrm{r} \leq 0.7$, follow a similar trend where the PM mechanical energy will redistribute between $f_\mathrm{TR}$ and $f_\mathrm{TR2}$ as $\lambda_\mathrm{TR}$ changes. However, a similar generic hypothesis for the interface and flow variables is difficult to make as they depend on other factors, e.g., the proximity of $f_\mathrm{TR},f_\mathrm{TR2}$, or $2f_\mathrm{TR}$ to the $f_\mathrm{VS}$.

In the interface variable plotted in Fig.~\ref{fig:FIG4}a.ii, the second truncation resonance remains the dominant spectral frequency for $\lambda_\mathrm{TR}=\lambda_1$. For $\lambda_\mathrm{TR}=\lambda_2$ three prominent spectral peaks at $f_\mathrm{TR}$, $f_\mathrm{TR2}$ and $2f_\mathrm{TR}$, with relatively decreasing magnitudes, in that order, are visible. For $\lambda_\mathrm{TR}\geq\lambda_3$, the $F_\mathrm{CS}$ spectral map is similar to the PM displacement (Fig.~\ref{fig:FIG4}a.i) with dominant spectral peaks close to $f_\mathrm{TR}$ and additional spectral peaks close to $2f_\mathrm{TR}$. Similar spectral characteristics are also visible in the flow variable (Fig.~\ref{fig:FIG4}a.iii) with a slight uptick in spectral energy at the second harmonic frequency, $2f_\mathrm{TR}$ observed for all $\{\lambda_{1},\cdots,\lambda_{6},\lambda_\mathrm{max}\}$, owing to global attractor effect.

For $f_\mathrm{r}=1.025$, the PM displacement response (Fig.~\ref{fig:FIG4}b.i) is highly narrow band at a shifted frequency, $f_\mathrm{TR}^\mathrm{FSI}$, that monotonically decreases with respect to the prescribed $f_\mathrm{TR}$, as $\lambda_\mathrm{TR}$ increases, (cf. the frequency shift observed in \citet{RamakrishnanJFS2026}). We hypothesize that this decrease in the fluid-PM coupling frequency, $f_{\mathrm{TR}}^\mathrm{FSI}$ ($<f_\mathrm{TR}$) results from a fluid-added mass effect. The stronger the fluid-PM coupling (i.e., higher the $\lambda_\mathrm{TR}$), higher the amount of fluid mass that latches on to the interface, altering the PM vibration characteristics and leading to a more pronounced decrease in $f_\mathrm{TR}^\mathrm{FSI}$. A quantitative relation linking the coupling strength, $\lambda_\mathrm{TR}$, to the coupling frequency, $f_\mathrm{TR}^\mathrm{FSI}$, is established later in Sec.~\ref{sec:Lambda_vs_fTR}.

Fig.~\ref{fig:FIG4}b.ii plots the frequency spectra of the interface variable, showing the presence of both the frequency shifted primary and secondary harmonics, $f_{\mathrm{TR}}^\mathrm{FSI}$ and $2f_{\mathrm{TR}}^\mathrm{FSI}$. Subsequently, in Fig.~\ref{fig:FIG4}b.iii that plots the spectra for the flow variable, only the primary harmonics survive owing to their proximity to $f_\mathrm{VS}$. Nevertheless, the frequency shift, evident across all the three PM-interface-flow parameters, indicates that the mechanism responsible for this frequency shift is necessarily a result of the fluid-PM coupling that simultaneously affects both the PM and the flow.

In summary, the PM dynamics undergo a shift in contribution of the truncation resonance modes as $\lambda_\mathrm{TR}$ changes. For low $\lambda_\mathrm{TR}$, the second truncation resonance remains localized at the FSI boundary like the primary truncation resonance, but with a high $\lambda_\mathrm{TR2}$. Therefore, when the frequency, $f_\mathrm{TR2}$ lies in the vicinity of the $f_\mathrm{VS}$, the second truncation mode is energetically preferred over the primary truncation mode in the PM dynamics. As $\lambda_\mathrm{TR}$ increases, the second truncation mode de-localizes from the FSI boundary and re-localizes at the opposite PM boundary decreasing $\lambda_\mathrm{TR2}$, making the primary truncation mode more energetically favorable. Therefore, regardless of the location of $f_\mathrm{TR,TR2}$ with respect to $f_\mathrm{VS}$, $f_\mathrm{TR}$ dominates the PM dynamics. This shift in PM dynamics from $f_\mathrm{TR2}$ to $f_\mathrm{TR}$, reflects in the flow and significantly contributes to distinct FSI regimes based on choice of $\lambda_\mathrm{TR}$.

\subsection{Distinct FSI regimes for different $f_\mathrm{TR}$ and $\lambda_\mathrm{TR}$}

We identify five distinct regimes of FSI, based on the choice of $\{f_\mathrm{TR},\lambda_\mathrm{TR}\}$:

\begin{enumerate}
    \item $f_\mathrm{r} \approx 0.5$, low $\lambda_\mathrm{TR}$: The PM displays multi-frequency dynamics with a broadband spectral distribution of energy across the primary and secondary truncation resonances, and the pass band frequencies, across the different PM-FSI cases. Specifically, the second truncation resonance emerges as a relatively dominant spectral frequency, when it approaches $f_\mathrm{VS}$. In the flow, in addition to the $f_\mathrm{TR,PB}$ frequencies, weak spectral signatures of the nonlinear second harmonic, $2f_\mathrm{TR}$, are also visible in the interface and flow variables, $\{F_\mathrm{CS},C_{l}\}$, rendering the spectral map to remain relatively broad band.
    \item $f_\mathrm{r} \approx 1$, low $\lambda_\mathrm{TR}$: The PM displays multi-frequency dynamics with a clear dominance of the primary truncation resonance, followed by a weak presence of the pass band frequencies, across the different PM-FSI cases. The flow spectral map resembles the PM spectral map with the addition of weak nonlinear second harmonics, generated at the interface ($F_\mathrm{CS}$). However, owing to these harmonics being far away from the global attractor frequency, $f_\mathrm{VS}$, their signatures disappear from the flow variable, $C_{l}$. 
    \item $f_\mathrm{r} \approx 0.5$, high $\lambda_\mathrm{TR}$: The PM displays narrow band dynamics, centered around the prescribed truncation resonance. $\{F_\mathrm{CS},C_{l}\}$, also display a high spectral energy concentration close to the $0.419 \leq f_\mathrm{r} \leq 0.513$, along with a strong presence of the second harmonic, near $f_\mathrm{VS}$ and activated by system nonlinearity.
    \item $f_\mathrm{r} \approx 1$, high $\lambda_\mathrm{TR}$: The FSI system, i.e., both the PM and the flow display narrow band dynamics, centered around the prescribed truncation resonance. $F_\mathrm{CS}$, displays strong nonlinear second harmonic generation. However, owing to these harmonics being far away from the global attractor frequency, $f_\mathrm{VS}$, their spectral signatures become relatively weaker in the flow variable, $C_{l}$, making the truncation resonance the dominant frequency to which the flow locks-on.
    \item $f_\mathrm{r} > 1$, any $\lambda_\mathrm{TR}$: The fluid-PM interface only features a static mean displacement with relatively muted subsurface dynamics. Consequently, the PM-FSI flow state approximately resembles the latent vortex-shedding state that was observed in the rigid plate case.
\end{enumerate}

\section{Dependence of FSI-induced vortex-shedding characteristics on the PM dynamics}\label{sec:EmpRel}
Having established the significance of the PM vibration amplitude envelope, $\lambda_\mathrm{TR}$, in determining the FSI dynamics, this section quantifies the effect of $\lambda_\mathrm{TR}$ on the fluid-PM coupling frequency, $f_\mathrm{TR}^\mathrm{FSI}$, and the lift force, $C_{l}$. Given the broad-band nature and participation of the second truncation resonance, $f_\mathrm{TR2}$, in the PM dynamics, for the low $\lambda_\mathrm{TR}$ cases with $0.419 \leq f_\mathrm{r} \leq 0.7$, in this section, we will primarily focus on the relation between $\lambda_\mathrm{TR}$ and $\{f_\mathrm{TR}^\mathrm{FSI},C_{l}\}$, only for PMs with $0.839 \leq f_\mathrm{r} \leq 1.4$. FSI simulations of PMs with $f_\mathrm{r}=2$ are also not considered here, owing to absence of significant FSI dynamics.

\begin{figure*}[t!]
    \centering
    \includegraphics[width=\textwidth]{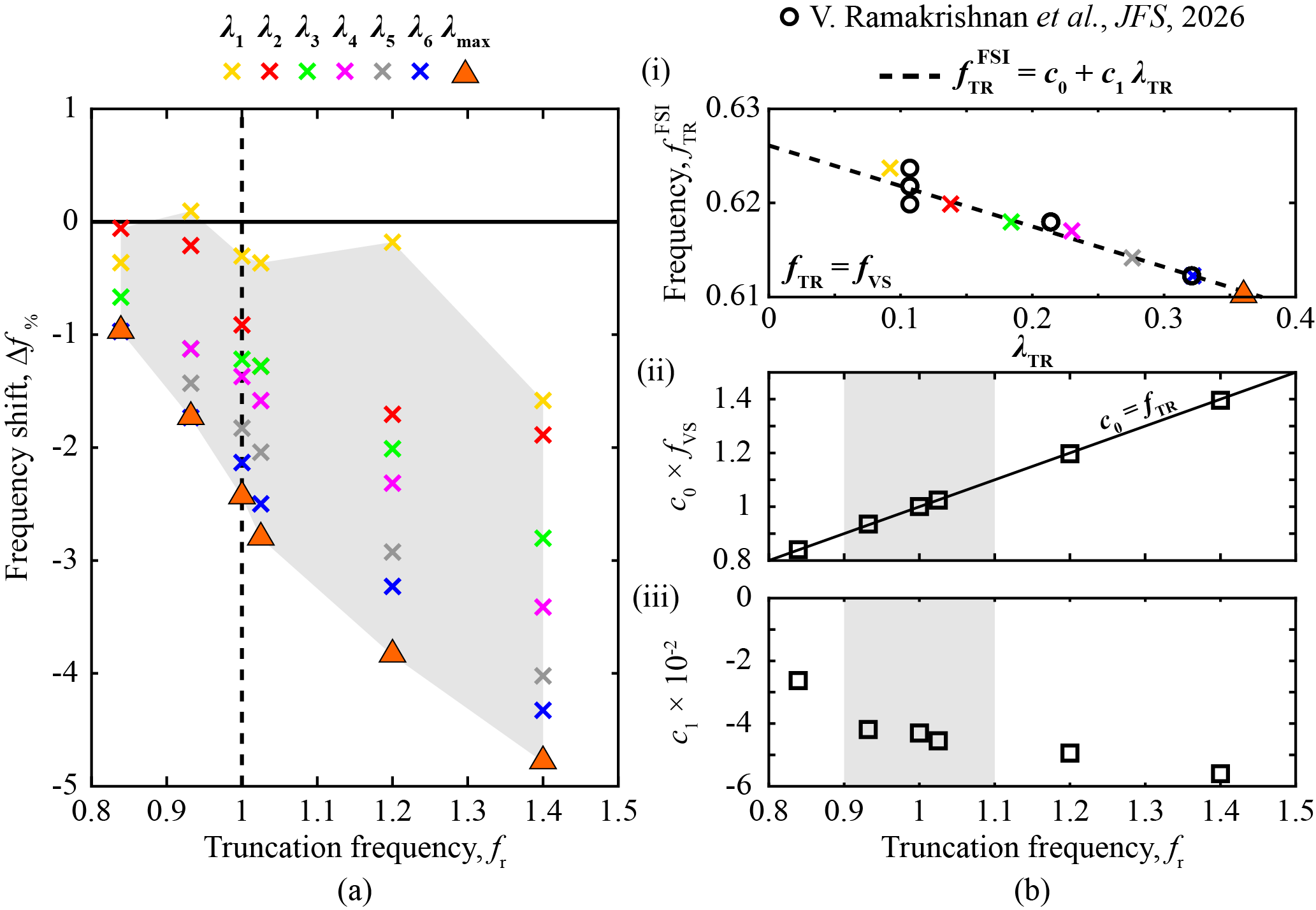}
    \caption{Dependence of the fluid-PM coupling frequency on the amplitude envelope. (a) Frequency shift, $\Delta f_{\%}$, observed in FSI simulations for PMs with $0.839 \leq f_\mathrm{r} \leq 2$. Shaded region depicts the range of $\Delta f_{\%}$ for different $f_\mathrm{TR}$. (b.i) $f_\mathrm{TR}^\mathrm{FSI}$ as a function of $\lambda_\mathrm{TR}$ for $f_\mathrm{r}=1$. (b.ii-iii) Linear fit parameters, $c_0$ and $c_1$, as functions of $f_\mathrm{r}$. Shaded region between $f_\mathrm{r}\in(0.9,1.1)$ denotes the region of best fit (high $R^2$ values) for the $\lambda_\mathrm{TR}$ and $f_\mathrm{TR}^\mathrm{FSI}$ data points.}
    \label{fig:FIG5}
\end{figure*}

\subsection{Effect of PM amplitude envelope on the fluid-PM coupling frequency}\label{sec:Lambda_vs_fTR}

In Sec.~\ref{sec:Results}, the Fourier analysis of FSI results revealed a monotonic decrease in the fluid-PM coupling frequency, $f_\mathrm{TR}^\mathrm{FSI}$ ($<f_\mathrm{TR}$), as the coupling strength, $\lambda_\mathrm{TR}$ increases. We now quantitatively analyze these FSI results to establish a relation between $\lambda_\mathrm{TR}$ and $f_\mathrm{TR}^\mathrm{FSI}$.

Fig.~\ref{fig:FIG5}a plots the frequency shift:
\begin{equation}
    \Delta f_{\%}=\left(\frac{f_\mathrm{TR}^\mathrm{FSI}-f_\mathrm{TR}}{f_\mathrm{TR}}\right) \times 100 \%
    \label{eq:Freq_Shift}
\end{equation}
observed in the displacement signal, $\chi$, in FSI simulations with $0.839 \leq f_\mathrm{r} \leq 1.4$, and $\{\lambda_\mathrm{TR}=\lambda_{1},\cdots,\lambda_{6},\lambda_\mathrm{max}\}$. Note that the same frequency shift in $f_\mathrm{TR}$ is also observed in the $\{F_\mathrm{CS},C_{l}\}$ signals (in addition to frequency shifts of the second harmonics of $f_\mathrm{TR}$ or the $f_\mathrm{TR2}$). We therefore focus only on the PM response for conciseness.

The shaded region in Fig.~\ref{fig:FIG5}a illustrates the range of frequency shift in the respective FSI simulations. The single mass-spring and the $\lambda_{1}$ PM-FSI cases result in the maximum and minimum frequency shifts, $|\Delta f_{\%}|$, respectively, for all $f_\mathrm{TR}$. The $\Delta f_{\%}$ values for the single mass-spring cases ($\Delta$ markers) display a linear increase in magnitude as the prescribed $f_\mathrm{TR}$ increases. Whereas, given the broad-band nature and weak fluid-PM coupling at $\lambda_{1}$ (gold $\mathbf{X}$ markers), we observe a lower $|\Delta f_{\%}|$ that does not follow a visible pattern as $f_\mathrm{TR}$ increases. However, as $\lambda_\mathrm{TR}$ approaches the single mass-spring limit (gray and blue $\mathbf{X}$ markers), the $|\Delta f_{\%}|$ linearly increases with $f_\mathrm{TR}$. Additionally, barring the minor discrepancies that may stem from the essentially non-linear fluid-PM interactions, for a given $f_\mathrm{TR}$, the $\mathbf{X}$ markers are approximately uniformly distributed in the $\Delta f_{\%}$ range, consistent with our choice of uniformly distributed $\lambda_{1,2,\cdots,6}$.

Fig.~\ref{fig:FIG5}b.i shows the relation between $\lambda_\mathrm{TR}$ and $f_\mathrm{TR}^\mathrm{FSI}$ for $f_\mathrm{r}=1$. The data is seen to align well with a best-fit linear curve,
\begin{equation}
    f_\mathrm{TR}^\mathrm{FSI}=c_0+c_1\lambda_\mathrm{TR},
    \label{eq:Lambda_fTR_EmpRel}
\end{equation}
using $c_0\approx f_\mathrm{TR}$ and $c_1=-0.043$. To further validate this relation, we also overlay the $f_\mathrm{r}=1$ data points from \citet{RamakrishnanJFS2026}, which explored FSI simulations at five different $m_\mathrm{UC}$ values and three distinct $\lambda_\mathrm{TR}$ (15 data points in total). Broadly, the linear fit suits the added data, though the $\mathbf{O}$ markers corresponding to the low $\lambda_\mathrm{TR}$ ($\approx0.107)$ cases spread out over a small range of $f_\mathrm{TR}^\mathrm{FSI}$. The markers for the intermediate $\lambda_\mathrm{TR}$ ($\approx0.214$) all coalesce, as do those for the high $\lambda_\mathrm{TR}$ ($\approx0.321$) cases. This outcome reinforces a previous deduction from \citet{RamakrishnanJFS2026}: $m_\mathrm{UC}$ has a reduced influence on FSI dynamics as the fluid-PM coupling $(\lambda_\mathrm{TR})$ grows stronger.

We now repeat this linear fitting exercise for all other FSI scenarios with $0.839 \leq f_\mathrm{r} \leq 1.4$ and obtain the fit parameters, $c_0$, and $c_1$. Fig.~\ref{fig:FIG5}b.ii shows that $c_0$ converges to the prescribed $f_\mathrm{TR}$ for all cases. This result is compatible with the FSI physics as $c_0$ represents the PM vibrational frequency, $f_\mathrm{TR}^\mathrm{FSI}\rightarrow f_\mathrm{TR}$, when $\lambda_\mathrm{TR}=0$. In other words, when the PM amplitude is negligible, there is no dynamic interaction with the aerodynamic flow and hence, there is no shift in the coupled frequency, i.e., $\left.f_\mathrm{TR}^\mathrm{FSI}\right|_{\lambda_\mathrm{TR}=0}=c_0=f_\mathrm{TR}$.

Fig.~\ref{fig:FIG5}b.iii plots the parameter $c_1$ for all the FSI cases. Similar to $|\Delta f_{\%}|$, a consistent increase in $|c_1|$ values are observed as $f_\mathrm{TR}$ increases. We hypothesize that this increase in both $|\Delta f_{\%}|$ and $|c_1|$ is a result of two competing effects in the FSI system. First, we have the global attractor effect in the fluid that favors the existence of frequencies that are proximal to $f_\mathrm{VS}$. Second, we have the fluid-added mass effect that tends to decrease the PM vibrational frequency as the fluid-PM coupling grows stronger. Therefore, for $f_\mathrm{r}>1$ both these effects work in consonance to decrease the $f_\mathrm{TR}^\mathrm{FSI}$, leading to higher $|\Delta f_{\%}|$ and $|c_1|$ values. Conversely, for $f_\mathrm{r}<1$ both effects counter each other, with the former trying to increase, and the latter trying to decrease $f_\mathrm{TR}^\mathrm{FSI}$, leading to relatively lower $|\Delta f_{\%}|$ and $|c_1|$ values.

Though, $|\Delta f_{\%}|$ and $|c_1|$ increase as $f_\mathrm{TR}$ increases, the rate of increase is slowest near $f_\mathrm{VS}$. We hypothesize this slower change to be an outcome of the flow locking onto the $f_\mathrm{TR}$ when that natural frequency is sufficiently near the latent vortex-shedding frequency. Therefore, the degree of lock-on and the frequency shift are both directly proportional to the fluid-PM coupling strength, $\lambda_\mathrm{TR}$. 

\begin{figure*}[t!]
    \centering
    \includegraphics[width=\textwidth]{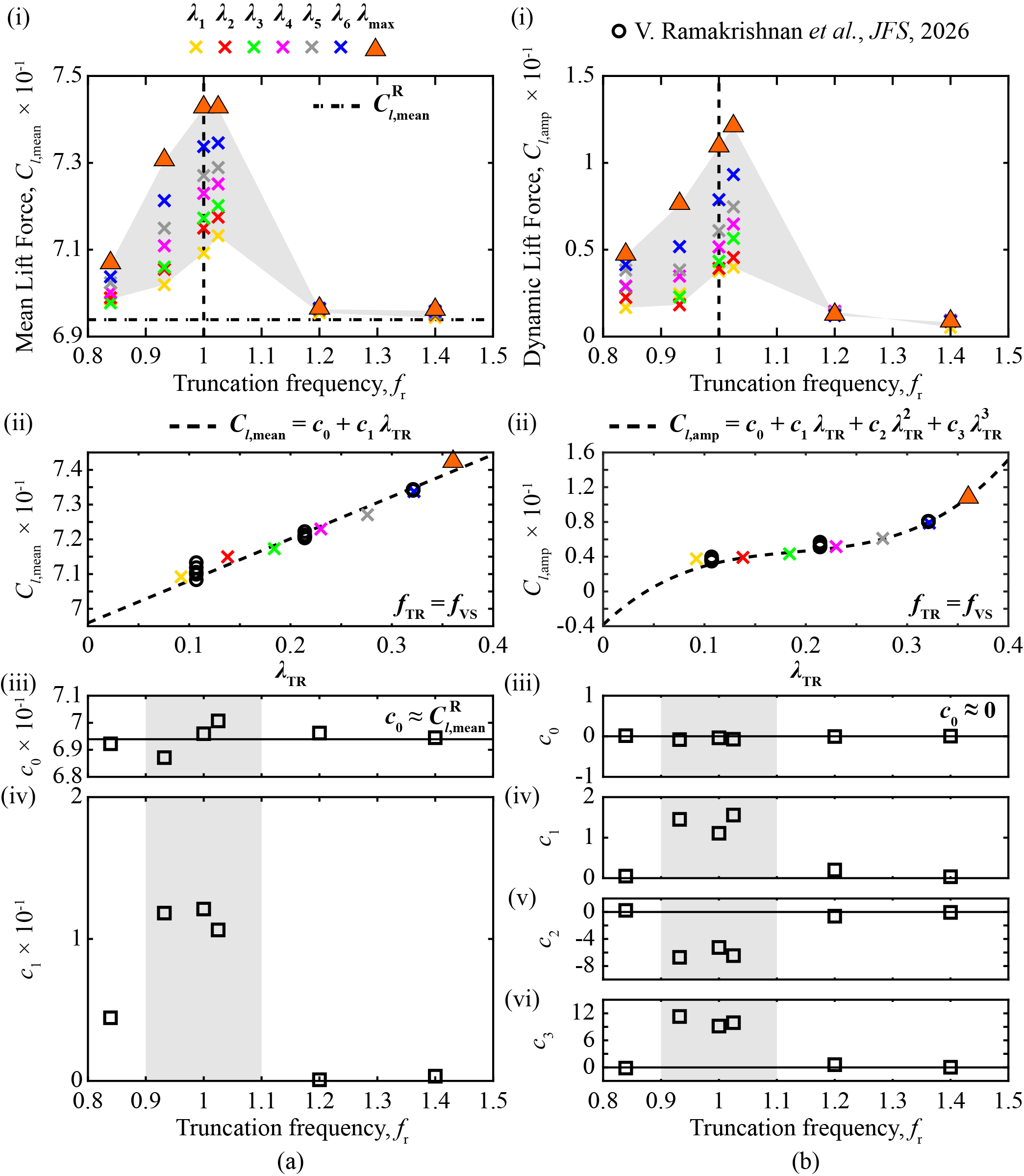}
    \caption{Dependence of the mean and dynamic components of the lift force on the amplitude envelope. (a,b.i) $C_{l,\mathrm{mean}}$ and $C_{l,\mathrm{amp}}$, observed in FSI simulations for PMs with $0.839 \leq f_\mathrm{r} \leq 1.4$. Shaded region depicts the range of $C_{l,\mathrm{mean}}$ and $C_{l,\mathrm{amp}}$ for different $f_\mathrm{TR}$. (a,b.ii) $C_{l,\mathrm{mean}}$ and $C_{l,\mathrm{amp}}$ as a function of $\lambda_\mathrm{TR}$ for $f_\mathrm{r}=1$. (a.iii-iv,b.iii-vi) The linear and cubic fit parameters, $c_{0,1}$, and $c_{0,1,2,3}$ for the mean and dynamic lift force components, as a function of $f_\mathrm{TR}$. Shaded region between $f_\mathrm{r}\in(0.9,1.1)$ denotes the region of best fit (high $R^2$ values) for the $\lambda_\mathrm{TR}$ and $C_{l,\mathrm{mean}}$, and $\lambda_\mathrm{TR}$ and $C_{l,\mathrm{amp}}$ data points.}\label{fig:FIG6}
\end{figure*}

\subsection{Effect of PM amplitude envelope on the lift force}\label{sec:Lambda_vs_Cl}

In this section, we study the effect of $\lambda_\mathrm{TR}$ on the lift force, $C_{l}$. Specifically, we separate the (static) mean and dynamic components of the lift force, i.e., $C_{l,\mathrm{mean}}$ and $C_{l,\mathrm{amp}}$, and individually analyze the effect of $\lambda_\mathrm{TR}$ on these components.

\subsubsection{Dependence of $C_{l,\mathrm{mean}}$ on $\lambda_\mathrm{TR}$}

Fig.~\ref{fig:FIG6}a.i plots the $C_{l,\mathrm{mean}}$ observed in FSI simulations with $0.839 \leq f_\mathrm{r} \leq 1.4$ and $\lambda_\mathrm{TR}\in\{\lambda_{1},\cdots,\lambda_{6},\lambda_\mathrm{max}\}$. The results reiterate the importance of strong PM dynamics $(\chi_\mathrm{amp}\neq0)$ to significantly alter the flow characteristics.  

The $C_{l,\mathrm{mean}}$ increases with increasing $\lambda_\mathrm{TR}$ that enables a stronger fluid-PM coupling. The single mass-spring and the $\lambda_{1}$ PM-FSI cases emerge as limiting cases defining the maximum and minimum $C_{l,\mathrm{mean}}$ values at a given $f_\mathrm{TR}$. The shaded region shows the range of the altered $C_{l,\mathrm{mean}}$, with the broadest range observed close to $f_\mathrm{r} \approx 1$. A maximum increase in the $C_{l,\mathrm{mean}}$ of $7 \%$ and $5.9 \%$, compared to the $C_{l,\mathrm{mean}}^\mathrm{R}$, is observed for the single mass-spring and the PM-FSI cases, respectively. Furthermore, since weak PM dynamics are observed for $f_\mathrm{r}>1$, the $C_{l,\mathrm{mean}}$ converges to the rigid place lift force ($C_{l,\mathrm{mean}}^\mathrm{R}$) in these FSI cases.

Fig.~\ref{fig:FIG6}a.ii shows the relation between $\lambda_\mathrm{TR}$ and $C_{l,\mathrm{mean}}$ data obtained from FSI simulations for $f_\mathrm{r}=1$. The data aligns with a best-fit linear curve,
\begin{equation}
    C_{l,\mathrm{mean}}=c_0+c_1\lambda_\mathrm{TR},
    \label{eq:Lambda_Cl_mean_EmpRel}
\end{equation}
with $c_0\approx C_{l,\mathrm{mean}}^\mathrm{R}$ and $c_1=0.121$. In addition, similar to Fig.~\ref{fig:FIG5}b.i, data points from \citet{RamakrishnanJFS2026} for FSI simulations with $f_\mathrm{r}=1$ are plotted as $\mathbf{O}$ markers. These data points also agree well with the linear fit.

Finally, Figs.~\ref{fig:FIG6}a.iii-iv plot the fit parameters for all other FSI scenarios with $0.839 \leq f_\mathrm{TR} \leq 1.4$. The $c_0$ value converges to the rigid plate mean lift force, $ C_{l,\mathrm{mean}}^\mathrm{R}$ for all the FSI cases. This result is similar to the results observed for the coupled frequency in Sec.~\ref{sec:Lambda_vs_fTR}. When $\lambda_\mathrm{TR}\approx0$, the lack of fluid-PM dynamic interaction causes the $C_{l,\mathrm{mean}}$ to converge to the rigid plate mean lift force, $C_{l,\mathrm{mean}}^\mathrm{R}$. The $c_1$ value remains approximately constant for PMs with $f_\mathrm{r}=0.932,1,1.025$ as these frequencies lie in the vicinity of $f_\mathrm{VS}$. A slightly lower $c_1$ value is observed for $f_\mathrm{r}=0.839$, and $c_1\approx0$ for $f_\mathrm{r}=1.2,1.4$. This indicates that strong coupled dynamics centered around coupled frequencies close to $f_\mathrm{VS}$ are most conducive to alter the mean lift force.

\subsubsection{Dependence of $C_{l,\mathrm{amp}}$ on $\lambda_\mathrm{TR}$}

Fig.~\ref{fig:FIG6}b.i plots the $C_{l,\mathrm{amp}}$ for FSI simulations with $0.839 \leq f_\mathrm{r} \leq 1.4$ and $\lambda_\mathrm{TR}=\{\lambda_{1},\cdots,\lambda_{6},\lambda_\mathrm{max}\}$. The range of $C_{l,\mathrm{amp}}$ (shaded region) qualitatively resembles the range of $C_{l,\mathrm{mean}}$. The value of $C_{l,\mathrm{amp}}\approx0$ for PMs with $f_\mathrm{r}=1.2,1.4$ re-iterate the presence of weak fluid-PM dynamics for all $\lambda_\mathrm{TR}$. Alternatively, the presence of significant PM dynamics for $0.839 \leq f_\mathrm{r} \lesssim 1$,  substantially alters the dynamic lift force, $C_{l,\mathrm{amp}}$.

The single mass-spring and the $\lambda_{1}$ FSI cases mark the extremes of the $C_{l,\mathrm{amp}}$ range. A maximum dynamic lift force, $C_{l,\mathrm{amp}}=0.121,0.093$, is obtained for the single mass-spring and PM-FSI cases, respectively. Interestingly, the $C_{l,\mathrm{amp}}$ attains its maximum value for PMs with $f_\mathrm{r}=1.025$, where the $f_\mathrm{TR}$ is slightly mis-aligned with the global attractor, $f_\mathrm{VS}$. The maximum $C_{l,\mathrm{mean}}$ observed for the PM-FSI occurs at the same frequency (blue $\mathbf{X}$ markers) in Fig.~\ref{fig:FIG6}a.i. We hypothesize this phenomena to be a result of the fluid-added mass effect that renders the coupling frequency, $f_\mathrm{TR}^\mathrm{FSI}\approx f_\mathrm{VS}$ for the $f_\mathrm{r}=1.025$ case. In other words, a slightly mis-aligned PM frequency relative to the latent vortex-shedding frequency accommodates a coupled frequency that is more closely aligned to the $f_\mathrm{VS}$, having a larger effect on the dynamic lift force.

Fig.~\ref{fig:FIG6}b.ii plots the $\lambda_\mathrm{TR}$ and $C_{l,\mathrm{amp}}$ simulation data points for PMs with $f_\mathrm{r}=1$. The data are well represented by a cubic polynomial fit,
\begin{equation}
C_{l,\mathrm{amp}}=c_0+c_1\lambda_\mathrm{TR}+c_2\lambda_\mathrm{TR}^2+c_3\lambda_\mathrm{TR}^3,
    \label{eq:Lambda_Cl_amp_EmpRel}
\end{equation}
with $c_0\approx 0$, $c_1=1.108$, $c_2=-5.251$ and $c_3=9.176$. The FSI data from our prior work~\citep{RamakrishnanJFS2026}, also agrees well with the cubic function.

Figs.~\ref{fig:FIG6}b.iii-vi plot the parameters $\{c_0,c_1,c_2,c_3\}$ for all the FSI scenarios with $0.839 \leq f_\mathrm{r} \leq 1.4$. The parameter $c_0\approx0$ for all FSI cases indicating that the dynamic lift oscillations don't exist in the absence of fluid-PM dynamic coupling (i.e., when $\lambda_\mathrm{TR}=0$). This observation is consistent with the lift force seen in the flow over a rigid plate, that also lacks any subsurface dynamics interacting with the flow ($C_{l,\mathrm{amp}}^\mathrm{R}=0$). For $f_\mathrm{r}=0.839,1.2,1.4$, where the truncation frequencies are away from $f_\mathrm{VS}$, we also have $c_{1,2,3}\approx 0$. For $f_\mathrm{r}\approx1$, the $c_{1,2,3}$ are approximately constant (and non-zero) due to the flow lock-on to the prescribed $f_\mathrm{TR}$ in these cases. This complements the earlier $\lambda_\mathrm{TR}$ and $C_{l,\mathrm{mean}}$ results, showing maximum flow modulations in FSI cases where the truncation resonance frequency lies in the vicinity latent vortex-shedding frequency ($f_\mathrm{r}\approx1$). Therefore, given that a steady flow exists in both the rigid plate (without any mean displacement or subsurface dynamics) and the PM-embedded flat plate (with nominal mean displacement and negligible subsurface dynamics) scenarios, we assert that the PM dynamics are the sole contributor to the existence of a dynamic component in the $C_{l}$.

\section{Conclusion}\label{sec:Conclusion}

Building off of the FSI-relevant behavioral parameters---the effective stiffness, $k_\mathrm{eff}$, unit cell mass, $m_\mathrm{UC}$, truncation resonance frequency, $f_\mathrm{TR}$, and the amplitude envelope, $\lambda_\mathrm{TR}$---and only a qualitative analysis of the FSI behavior over a limited range of behavioral parameters, this study performs an exhaustive sweep of the key behavioral parameters, establishes scaling relations between behavioral parameters and the coupled behavior, and identifies distinct FSI regimes when varying these parameters. We identify $f_\mathrm{TR}$ and $\lambda_\mathrm{TR}$ as the two most important PM behavioral parameters that dictate the FSI, and quantify the effect of these parameters on the output quantities such as coupled frequency, the lift force, and circulation.

Three distinct regimes of FSI behaviors emerge based on the choice of the truncation resonance: $f_\mathrm{r}\approx0.5$, $f_\mathrm{r}\approx1$, and $f_\mathrm{r}>1$. In addition, two distinct sub-regimes for each of the $f_\mathrm{r}\approx0.5$ and $f_\mathrm{r}\approx1$ cases are observed based on the choice of the amplitude envelope (low or high $\lambda_\mathrm{TR}\in(0,\lambda_\mathrm{max})$), establishing a total of five distinct regimes of FSI. Briefly, for $f_\mathrm{r}\approx0.5$, multi-frequency FSI dynamics are observed with significant participation from the primary truncation resonance, the second truncation resonance and the nonlinear second harmonic of the primary truncation resonance. Consequently, given the broadband and shift in dominant PM resonance frequency, the $f_\mathrm{TR}$ and $\lambda_\mathrm{TR}$ cannot fully capture the FSI dynamics in this regime. For $f_\mathrm{r}\approx1$, the behavioral parameters dictate the narrow band FSI dynamics, where the fluid-PM system locks onto a frequency slightly lower than the prescribed $f_\mathrm{TR}$. For $f_\mathrm{r}>1$, negligible coupled dynamics are observed that causes the perturbed flow to the latent vortex-shedding state as seen in the rigid plate case.

When $f_\mathrm{r}\approx 0.5$, the coupled dynamics are broadband and are dominated by the second truncation resonance or the second harmonic of the primary truncation resonance for low $\lambda_\mathrm{TR}$. As $\lambda_\mathrm{TR}$ increases, the coupled dynamics feature strong spectral peaks at the primary truncation resonance frequency and the corresponding second harmonic. Since multiple modes dominate the FSI as opposed to the designed truncation mode, a clear relation between the important behavioral parameters and the coupled parameters cannot be established. Alternatively, when $f_\mathrm{r}\approx 1$, the flow locks-on close to the prescribed truncation resonance frequency, showing explicit dependence of the coupled FSI quantities, $\{f_\mathrm{TR}^\mathrm{FSI},C_{l}\}$ on $\lambda_\mathrm{TR}$. The coupled frequency, $f_\mathrm{TR}^\mathrm{FSI}$ is \textit{linearly} dependent on the FSI coupling strength, $\lambda_\mathrm{TR}$. The mean and dynamic lift force, $C_{l,\mathrm{mean}}$ and $C_{l,\mathrm{amp}}$, respectively, have a \textit{linear} and \textit{cubic} relation to $\lambda_\mathrm{TR}$. We also observe that as the strength of fluid-PM coupling (i.e., $\lambda_\mathrm{TR}$) increases, frequency signatures of non-linear super harmonics arise (strong signatures in $F_\mathrm{CS}$, and relatively weaker in $C_{l}$) and the coupled frequency, $f_\mathrm{TR}^\mathrm{FSI}$ decreases relative to $f_\mathrm{TR}$ due to a fluid-added mass effect. These results comprehensively show the PM ability to dictate the strength, and the spatio-temporal spectral characteristics of the FSI-induced vortex-shedding process.

In conclusion, this study comprehensively analyzes the PM-FSI between a grounded diatomic phononic material and vortex-shedding behavior past an inclined flat plate. Though this study focused on FSI in a canonical aerodynamic flow, adapting the behavioral parameter-based PM framework~\citep{RamakrishnanJFS2026} and the quantitative relations between the PM behavioral parameters and the coupled output parameters, could be a pathway towards designing PMs for FSI and flow control applications.
    
\section*{CRediT authorship contribution statement}
Vinod Ramakrishnan (VR): investigation, formal analysis, data curation, visualization, writing – original draft, review, and editing. Arturo Burgos (AB): investigation, formal analysis. Sangwon Park (SP): investigation, formal analysis. Andres Goza (AG): conceptualization, supervision, formal analysis, funding acquisition, writing–review, and editing. Kathryn Matlack (KM): conceptualization, supervision, formal analysis, funding acquisition, writing–review, and editing. 

\section*{Data availability} 
Data will be made available on request. 

\section*{Acknowledgments} 
This material is based upon work supported by the Air Force Office of Scientific Research under award number FA9550-21-1-0182 and award number FA9550-23-1-0299, and the Grainger College of Engineering at the University of Illinois Urbana-Champaign through the Strategic Research Initiative.

\newpage

\section*{Appendix}

\setcounter{equation}{0}
\renewcommand{\theequation}{A\arabic{equation}}
\renewcommand{\theHequation}{A\arabic{equation}}
\setcounter{figure}{0}
\renewcommand{\thefigure}{A\arabic{figure}}
\renewcommand{\theHfigure}{A\arabic{figure}}
\setcounter{section}{0}
\renewcommand{\thesection}{A\arabic{section}}
\renewcommand{\theHsection}{A\arabic{section}}
\setcounter{table}{0}
\renewcommand{\thetable}{A\arabic{table}}
\renewcommand{\theHtable}{A\arabic{table}}

\section{PM Structural Parameters for FSI Simulations}

Tab.~\ref{tab:FSI_PM_Sims} provides the 78 distinct PM structural and behavioral parameters explored for FSI simulations in this article. Note that all PMs have an identical effective stiffness, $k_\mathrm{eff}=5.4533$.

\renewcommand{\arraystretch}{1.5}
\begin{table}[H]
    \centering
    \begin{tabular}{|M{2cm}|M{2.3cm}|M{2.3cm}|M{2.3cm}|M{2.3cm}|M{2.3cm}|}\hline
    \multicolumn{6}{|c|}{$\bm{f_\mathrm{TR}/f_\mathrm{VS}=f_\mathrm{r}, f_\mathrm{VS}=0.6256,\{m_1,m_2,k,k_\mathrm{g}\}}$}\\ \hline    $\bm{\left[f_\mathrm{r},m_\mathrm{UC},\lambda_0\right]}$ & $\bm{\left[0.419,\right.}$ $\bm{\left. 8.889,0.135\right]}$ & $\bm{\left[0.466,\right.}$ $\bm{\left. 7.2,0.15\right]}$ & $\bm{\left[0.5,\right.}$ $\bm{\left. 6.253,0.161\right]}$ & $\bm{\left[0.513,\right.}$ $\bm{\left. 5.95,0.165\right]}$ & $\bm{\left[0.6,\right.}$ $\bm{\left. 4.342,0.193\right]}$ \\ \hline
    $\bm{\lambda_1\left(=\frac{2\lambda_0}{7}\right)}$ & $\{0.83,8.059$, $6.76,25.282\}$ & $\{0.672,6.529$, $6.76,25.282\}$ & $\{0.583,5.67$, $6.76,25.282\}$ & $\{0.555,5.396$, $6.76,25.282\}$ & $\{0.405,3.938$, $6.76,25.282\}$\\ \hline
    $\bm{\lambda_2\left(=\frac{3\lambda_0}{7}\right)}$ & $\{1.072,7.818$, $6.7,26.398\}$ & $\{0.869,6.332$, $6.7,26.398\}$ & $\{0.754,5.499$, $6.7,26.398\}$ & $\{0.718,5.233$, $6.7,26.398\}$ & $\{0.524,3.819$, $6.7,26.398\}$\\ \hline
    $\bm{\lambda_3\left(=\frac{4\lambda_0}{7}\right)}$ & $\{1.264,7.624$, $6.63,27.846\}$ & $\{1.024,6.175$, $6.63,27.846\}$ & $\{0.889,5.363$, $6.63,27.846\}$ & $\{0.847,5.102$, $6.62,27.87\}$ & $\{0.618,3.723$, $6.62,27.87\}$\\ \hline
    $\bm{\lambda_4\left(=\frac{5\lambda_0}{7}\right)}$ & $\{1.438,7.45$, $6.54,29.953\}$ & $\{1.165,6.034$, $6.53,29.973\}$ & $\{1.012,5.24$, $6.53,29.973\}$  & $\{0.963,4.987$, $6.53,29.973\}$ & $\{0.703,3.639$, $6.53,29.973\}$\\ \hline
    $\bm{\lambda_5\left(=\frac{6\lambda_0}{7}\right)}$ & $\{1.607,7.282$, $6.42,33.448\}$ & $\{1.302,5.898$, $6.42,33.448\}$ & $\{1.131,5.122$, $6.42,33.448\}$ & $\{1.076,4.874$, $6.42,33.448\}$ & $\{0.785,3.557$, $6.42,33.448\}$\\ \hline
    $\bm{\lambda_6\left(=\lambda_0\right)}$ & $\{1.767,7.122$, $6.24,40.186\}$ & $\{1.431,5.769$, $6.24,40.186\}$ & $\{1.253,4.999$, $6.23,40.993\}$ & $\{1.192,4.757$, $6.23,40.993\}$ & $\{0.87,3.472$, $6.23,40.993\}$\\ \hline
    \end{tabular}\\[0.2cm]
    \begin{tabular}{|M{2cm}|M{2.3cm}|M{2.3cm}|M{2.3cm}|M{2.3cm}|M{2.3cm}|}\hline
    $\bm{\left[f_\mathrm{r},m_\mathrm{UC},\lambda_0\right]}$ & $\bm{\left[0.7,\right.}$ $\bm{\left. 3.19,0.225\right]}$ & $\bm{\left[0.839,\right.}$ $\bm{\left. 2.222,0.27\right]}$ & $\bm{\left. 0.932,\right.}$ $\bm{\left. 1.8,0.3\right]}$ & $\bm{\left[1,\right.}$ $\bm{\left. 1.563,0.322\right]}$ & $\bm{\left[1.025,\right.}$ $\bm{\left. 1.488,0.33\right]}$\\ \hline
    $\bm{\lambda_1\left(=\frac{2\lambda_0}{7}\right)}$ & $\{0.297,2.894$, $6.76,25.282\}$ & $\{0.207,2.016$, $6.76,25.282\}$ & $\{0.167,1.634$, $6.76,25.282\}$ & $\{0.145,1.419$, $6.76,25.282\}$ & $\{0.138,1.35$, $6.76,25.282\}$\\ \hline
    $\bm{\lambda_2\left(=\frac{3\lambda_0}{7}\right)}$ & $\{0.385,2.806$, $6.7,26.398\}$ & $\{0.268,1.955$, $6.7,26.398\}$ & $\{0.217,1.584$, $6.7,26.398\}$ & $\{0.188,1.376$, $6.7,26.398\}$ & $\{0.179,1.309$, $6.7,26.398\}$\\ \hline
    $\bm{\lambda_3\left(=\frac{4\lambda_0}{7}\right)}$ & $\{0.454,2.735$, $6.62,27.87\}$ & $\{0.317,1.905$, $6.62,27.87\}$ & $\{0.256,1.543$, $6.62,27.87\}$ & $\{0.223,1.34$, $6.62,27.87\}$ & $\{0.212,1.275$, $6.62,27.87\}$\\ \hline
    $\bm{\lambda_4\left(=\frac{5\lambda_0}{7}\right)}$ & $\{0.516,2.674$, $6.53,29.973\}$ & $\{0.36,1.862$, $6.53,29.973\}$ & $\{0.291,1.509$, $6.53,29.973\}$ & $\{0.253,1.31$, $6.53,29.973\}$ & $\{0.241,1.247$, $6.53,29.973\}$\\ \hline
    $\bm{\lambda_5\left(=\frac{6\lambda_0}{7}\right)}$ & $\{0.577,2.613$, $6.42,33.448\}$ & $\{0.402,1.82$, $6.42,33.448\}$ & $\{0.326,1.475$, $6.42,33.448\}$ & $\{0.283,1.281$, $6.42,33.448\}$ & $\{0.269,1.219$, $6.42,33.448\}$\\ \hline
    $\bm{\lambda_6\left(=\lambda_0\right)}$ & $\{0.639,2.551$, $6.23,40.993\}$ & $\{0.445,1.777$, $6.23,40.993\}$ & $\{0.361,1.439$, $6.23,40.993\}$ & $\{0.313,1.25$, $6.23,40.993\}$ & $\{0.298,1.189$, $6.23,40.993\}$\\ \hline
    \end{tabular}\\[0.2cm]
\end{table}
\begin{table}[H]
    \centering
    \begin{tabular}{|M{2cm}|M{2.3cm}|M{2.3cm}|M{2.3cm}|}\hline  $\bm{\left[f_\mathrm{r},m_\mathrm{UC},\lambda_0\right]}$ & $\bm{\left[1.2,\right.}$ $\bm{\left. 1.086,0.386\right]}$ & $\bm{\left[1.4,\right.}$ $\bm{\left. 0.798,0.451\right]}$ & $\bm{\left[2,\right.}$ $\bm{\left. 0.391,0.644\right]}$\\ \hline
    $\bm{\lambda_1\left(=\frac{2\lambda_0}{7}\right)}$ & $\{0.101,0.986$, $6.76,25.282\}$ & $\{0.074,0.724$, $6.76,25.282\}$ & $\{0.0362,0.354$, $6.76,25.215\}$\\ \hline
    $\bm{\lambda_2\left(=\frac{3\lambda_0}{7}\right)}$ & $\{0.131,0.956$, $6.7,26.398\}$ & $\{0.096,0.702$, $6.7,26.398\}$ & $\{0.047,0.344$, $6.7,26.398\}$\\ \hline
    $\bm{\lambda_3\left(=\frac{4\lambda_0}{7}\right)}$ & $\{0.155,0.931$, $6.62,27.87\}$ & $\{0.114,0.683$, $6.63,27.846\}$ & $\{0.056,0.336$, $6.62,27.936\}$\\ \hline
    $\bm{\lambda_4\left(=\frac{5\lambda_0}{7}\right)}$ & $\{0.176,0.91$, $6.53,29.973\}$ & $\{0.129,0.668$, $6.53,29.973\}$ & $\{0.063,0.327$, $6.54,29.953\}$\\ \hline
    $\bm{\lambda_5\left(=\frac{6\lambda_0}{7}\right)}$ & $\{0.196,0.889$, $6.42,33.448\}$ & $\{0.144,0.653$, $6.42,33.384\}$ & $\{0.071,0.32$, $6.42,33.384\}$\\ \hline
    $\bm{\lambda_6\left(=\lambda_0\right)}$ & $\{0.218,0.868$, $6.23,40.993\}$ & $\{0.16,0.638$, $6.23,40.993\}$ & $\{0.078,0.312$, $6.23,40.993\}$\\ \hline
    \end{tabular}
    \caption{FSI PM behavioral and structural parameters. ($k_\mathrm{eff}=5.4533$)}
    \label{tab:FSI_PM_Sims}
\end{table}

Tab.~\ref{tab:FSI_SDOF_Sims} provides the 13 distinct single mass-spring structural and behavioral parameters explored for FSI simulations in this article.

\begin{table}[H]
    \centering
    \begin{tabular}{|M{0.3cm}|M{1.9cm}|M{1.9cm}|M{1.9cm}|M{1.9cm}|M{1.9cm}|M{1.9cm}|M{1.9cm}|}\hline
    \multicolumn{8}{|c|}{$\bm{f_\mathrm{s}/f_\mathrm{VS}=f_\mathrm{r}, f_\mathrm{VS}=0.6256,\{m_\mathrm{s},\lambda_\mathrm{max}\}}$}\\ \hline
    $\bm{f_\mathrm{r}}$ & $\bm{0.419}$ & $\bm{0.466}$ & $\bm{0.5}$ & $\bm{0.513}$ & $\bm{0.6}$ & $\bm{0.7}$ & $\bm{0.839}$\\ \hline
     & $\{2.007,0.151\}$ & $\{1.626,0.168\}$ & $\{1.412,0.18\}$ & $\{1.343,0.185\}$ & $\{0.98,0.216\}$ & $\{0.72,0.252\}$ & $\{0.502,0.302\}$\\ \hline
    \end{tabular}\\[0.2cm]
    \begin{tabular}{|M{0.3cm}|M{1.9cm}|M{1.9cm}|M{1.9cm}|M{1.9cm}|M{1.9cm}|M{1.9cm}|}\hline
    $\bm{f_\mathrm{r}}$ & $\bm{0.932}$ & $\bm{1}$ & $\bm{1.025}$ & $\bm{1.2}$ & $\bm{1.4}$ & $\bm{2}$\\ \hline
    & $\{0.406,0.336\}$ & $\{0.353,0.36\}$ & $\{0.336,0.369\}$ & $\{0.245,0.432\}$ & $\{0.18,0.505\}$ & $\{0.088,0.721\}$\\ \hline
    \end{tabular}
    \caption{FSI Single mass-spring behavioral and structural parameters. ($k_\mathrm{s}=k_\mathrm{eff}=5.4533$)}
    \label{tab:FSI_SDOF_Sims}
\end{table}

We can draw a few observations from the above parameter sets:
\begin{itemize}
    \item The unit cell mass, $m_\mathrm{UC}$, scales with $1/f_\mathrm{TR}^2$, consistent with $f\propto1/\sqrt{m}$ for diatomic PMs.
    \item The displacement envelopes, $\lambda_\mathrm{max}$, and $\lambda_0$ (and consequently $\lambda_{1,2,\cdots,6}$) scale with $f_\mathrm{TR}$, consistent with PM velocity $\propto f$ ($\lambda$ has units of velocity).
    \item The mass ratio, $m_\mathrm{r}=m_2/m_1$, remains constant for a proportional $\{\lambda,m_\mathrm{UC}\}$, across different $f_\mathrm{TR}$, with only the absolute mass changing to produce the desirable PM behaviors. For example, consider $\{\lambda_4,m_\mathrm{UC}\}$ for all $f_\mathrm{r}\in[0.419,2]$ : $ \ m_\mathrm{r}=\frac{7.45}{1.438}=\frac{6.034}{1.165}=\frac{5.24}{1.012}=\frac{4.987}{0.963}=\frac{3.639}{0.703}=\frac{2.674}{0.516}=\frac{1.862}{0.36}=\frac{1.509}{0.291}=\frac{1.31}{0.253}=\frac{1.247}{0.241}=\frac{0.91}{0.176}=\frac{0.668}{0.129}=\frac{0.327}{0.063}\approx5.18$.
    \item The absolute stiffnesses, $k_\mathrm{g}$, and $k$, and consequently, the grounding stiffness ratio, $k_\mathrm{gr}$, both remain constant for a given $\{\lambda,m_\mathrm{UC}\}$ across different $f_\mathrm{TR}$. For example, consider $\{\lambda_4,m_\mathrm{UC}\}$ for all $f_\mathrm{r}\in[0.419,2]$ : $\{k,k_\mathrm{g}\}=\{6.7,26.398\}$.
\end{itemize}

Tab.~\ref{tab:PM_TR2_Modes} provides the PM structural for behavioral parameters: $\{k_\mathrm{eff},m_\mathrm{UC},f_\mathrm{TR},\lambda_\mathrm{TR}\}=\{5.4533,1.563,0.5f_\mathrm{VS}$, $[\lambda_{1},\cdots,\lambda_{6}]\}$, used to calculate the $f_\mathrm{TR2}$ resonance modes illustrated in Fig.~\ref{fig:FIG3}c.

\begin{table}[H]
    \centering
    \begin{tabular}{|M{2.4cm}|M{1.9cm}|M{1.9cm}|M{1.9cm}|M{1.9cm}|M{1.9cm}|M{1.9cm}|}\hline
    & $\bm{\lambda_1\left(=\frac{2\lambda_0}{7}\right)}$ & $\bm{\lambda_2\left(=\frac{3\lambda_0}{7}\right)}$ & $\bm{\lambda_3\left(=\frac{4\lambda_0}{7}\right)}$ & $\bm{\lambda_4\left(=\frac{5\lambda_0}{7}\right)}$ & $\bm{\lambda_5\left(=\frac{6\lambda_0}{7}\right)}$ & $\bm{\lambda_6\left(=\lambda_0\right)}$\\ \hline
    $\bm{\left[f_\mathrm{r},m_\mathrm{UC},\lambda_0\right]}$ = $[0.5,1.563,0.162]$ & $\{0.374,1.19$, $12,6.24\}$ & $\{0.558,1.001$, $11.66,6.53\}$ & $\{0.717,0.846$, $11.21,6.95\}$ & $\{0.878,0.685$, $10.6,7.632\}$ & $\{1.056,0.507$, $9.68,9.002\}$ & $\{1.25,0.312$, $8.22,12.905\}$\\ \hline
    \end{tabular}
    \caption{PM behavioral and structural parameters corresponding to $f_\mathrm{TR2}$ modes in Fig.~\ref{fig:FIG3}c. ($k_\mathrm{eff}=5.4533$)}
    \label{tab:PM_TR2_Modes}
\end{table}

\newpage

\begin{thebibliography}{30}
\expandafter\ifx\csname natexlab\endcsname\relax\def\natexlab#1{#1}\fi
\providecommand{\url}[1]{\texttt{#1}}
\providecommand{\href}[2]{#2}
\providecommand{\path}[1]{#1}
\providecommand{\DOIprefix}{doi:}
\providecommand{\ArXivprefix}{arXiv:}
\providecommand{\URLprefix}{URL: }
\providecommand{\Pubmedprefix}{pmid:}
\providecommand{\doi}[1]{\href{http://dx.doi.org/#1}{\path{#1}}}
\providecommand{\Pubmed}[1]{\href{pmid:#1}{\path{#1}}}
\providecommand{\bibinfo}[2]{#2}
\ifx\xfnm\relax \def\xfnm[#1]{\unskip,\space#1}\fi
\bibitem[{Kushwaha et~al.(1993)Kushwaha, Halevi, Dobrzynski, and Djafari-Rouhani}]{KushwahaPRL1993}
\bibinfo{author}{M.~S. Kushwaha}, \bibinfo{author}{P.~Halevi}, \bibinfo{author}{L.~Dobrzynski}, \bibinfo{author}{B.~Djafari-Rouhani},
\newblock \bibinfo{title}{Acoustic band structure of periodic elastic composites},
\newblock \bibinfo{journal}{Phys. Rev. Lett.} \bibinfo{volume}{71} (\bibinfo{year}{1993}) \bibinfo{pages}{2022}.
\bibitem[{Bastawrous and Hussein(2022)}]{BastawrousJASA2022}
\bibinfo{author}{M.~V. Bastawrous}, \bibinfo{author}{M.~I. Hussein},
\newblock \bibinfo{title}{Closed-form existence conditions for bandgap resonances in a finite periodic chain under general boundary conditions},
\newblock \bibinfo{journal}{J. Acoust. Soc. Am.} \bibinfo{volume}{151} (\bibinfo{year}{2022}) \bibinfo{pages}{286--298}.
\bibitem[{Al~Ba'ba'a et~al.(2019)Al~Ba'ba'a, Nouh, and Singh}]{HasanPRSA2019}
\bibinfo{author}{H.~Al~Ba'ba'a}, \bibinfo{author}{M.~Nouh}, \bibinfo{author}{T.~Singh},
\newblock \bibinfo{title}{Dispersion and topological characteristics of permutative polyatomic phononic crystals},
\newblock \bibinfo{journal}{Proc. R. Soc. A} \bibinfo{volume}{475} (\bibinfo{year}{2019}) \bibinfo{pages}{20190022}.
\bibitem[{Al~Ba'ba'a et~al.(2024)Al~Ba'ba'a, Yousef, and Nouh}]{HasanJEL2024}
\bibinfo{author}{H.~B. Al~Ba'ba'a}, \bibinfo{author}{H.~Yousef}, \bibinfo{author}{M.~Nouh},
\newblock \bibinfo{title}{{A blueprint for truncation resonance placement in elastic diatomic lattices with unit cell asymmetry}},
\newblock \bibinfo{journal}{JASA Express Lett.} \bibinfo{volume}{4} (\bibinfo{year}{2024}) \bibinfo{pages}{077501}.
\bibitem[{Ramakrishnan and Matlack(2025)}]{RamakrishnanJSV2025}
\bibinfo{author}{V.~Ramakrishnan}, \bibinfo{author}{K.~H. Matlack},
\newblock \bibinfo{title}{A quantitative study of energy localization characteristics in defect-embedded monoatomic phononic crystals},
\newblock \bibinfo{journal}{J. Sound Vib.} \bibinfo{volume}{614} (\bibinfo{year}{2025}) \bibinfo{pages}{119164}.
\bibitem[{Lee and Kim(2023)}]{LeeSMS2023}
\bibinfo{author}{J.~Lee}, \bibinfo{author}{Y.~Y. Kim},
\newblock \bibinfo{title}{Elastic metamaterials for guided waves: from fundamentals to applications},
\newblock \bibinfo{journal}{Smart Mater. Struct.} \bibinfo{volume}{32} (\bibinfo{year}{2023}) \bibinfo{pages}{123001}.
\bibitem[{Miniaci and Pal(2021)}]{MiniaciJAP2021}
\bibinfo{author}{M.~Miniaci}, \bibinfo{author}{R.~K. Pal},
\newblock \bibinfo{title}{{Design of topological elastic waveguides}},
\newblock \bibinfo{journal}{J. Appl. Phys.} \bibinfo{volume}{130} (\bibinfo{year}{2021}) \bibinfo{pages}{141101}.
\bibitem[{Kushwaha(1996)}]{KushwahaIJMPB1996}
\bibinfo{author}{M.~S. Kushwaha},
\newblock \bibinfo{title}{Classical band structure of periodic elastic composites},
\newblock \bibinfo{journal}{Int. J. Mod. Phys. B} \bibinfo{volume}{10} (\bibinfo{year}{1996}) \bibinfo{pages}{977--1094}.
\bibitem[{Jo and Youn(2022)}]{JoMAMS2022}
\bibinfo{author}{S.-H. Jo}, \bibinfo{author}{B.~D. Youn},
\newblock \bibinfo{title}{Designing a phononic crystal with a defect for target frequency matching using an analytical approach},
\newblock \bibinfo{journal}{Mech. Adv. Mater. Struct.} \bibinfo{volume}{29} (\bibinfo{year}{2022}) \bibinfo{pages}{2454--2467}.
\bibitem[{Schmidt et~al.(2025)Schmidt, Yousef, Roy, Scalo, and Nouh}]{SchmidtJAP2025}
\bibinfo{author}{R.~Schmidt}, \bibinfo{author}{H.~Yousef}, \bibinfo{author}{I.~Roy}, \bibinfo{author}{C.~Scalo}, \bibinfo{author}{M.~Nouh},
\newblock \bibinfo{title}{Perturbation energy extraction from a fluid via a subsurface acoustic diode with sustained downstream attenuation},
\newblock \bibinfo{journal}{J. Appl. Phys.} \bibinfo{volume}{137} (\bibinfo{year}{2025}).
\bibitem[{Hussein et~al.(2015)Hussein, Biringen, Bilal, and Kucala}]{hussein2015flow}
\bibinfo{author}{M.~I. Hussein}, \bibinfo{author}{S.~Biringen}, \bibinfo{author}{O.~R. Bilal}, \bibinfo{author}{A.~Kucala},
\newblock \bibinfo{title}{Flow stabilization by subsurface phonons},
\newblock \bibinfo{journal}{Proc. R. Soc. A} \bibinfo{volume}{471} (\bibinfo{year}{2015}) \bibinfo{pages}{20140928}.
\bibitem[{Willey et~al.(2023)Willey, Barnes, Chen, Rosenberg, Medina, and Juhl}]{WilleyJFS23}
\bibinfo{author}{C.~L. Willey}, \bibinfo{author}{C.~J. Barnes}, \bibinfo{author}{V.~W. Chen}, \bibinfo{author}{K.~Rosenberg}, \bibinfo{author}{A.~Medina}, \bibinfo{author}{A.~T. Juhl},
\newblock \bibinfo{title}{Multi-input multi-output phononic subsurfaces for passive boundary layer transition delay},
\newblock \bibinfo{journal}{J. Fluids Struct.} \bibinfo{volume}{121} (\bibinfo{year}{2023}) \bibinfo{pages}{103936}.
\bibitem[{Michelis et~al.(2023)Michelis, Putranto, and Kotsonis}]{MichelisPoF2023}
\bibinfo{author}{T.~Michelis}, \bibinfo{author}{A.~Putranto}, \bibinfo{author}{M.~Kotsonis},
\newblock \bibinfo{title}{Attenuation of tollmien--schlichting waves using resonating surface-embedded phononic crystals},
\newblock \bibinfo{journal}{Phys. Fluids} \bibinfo{volume}{35} (\bibinfo{year}{2023}).
\bibitem[{Barnes et~al.(2021)Barnes, Willey, Rosenberg, Medina, and Juhl}]{BarnesAIAA2021}
\bibinfo{author}{C.~J. Barnes}, \bibinfo{author}{C.~L. Willey}, \bibinfo{author}{K.~Rosenberg}, \bibinfo{author}{A.~Medina}, \bibinfo{author}{A.~T. Juhl},
\newblock \bibinfo{title}{Initial computational investigation toward passive transition delay using a phononic subsurface},
\newblock in: \bibinfo{booktitle}{AIAA Scitech 2021 Forum}, \bibinfo{year}{2021}, p. \bibinfo{pages}{1454}.
\bibitem[{Keogh et~al.(2025)Keogh, McTighe, Dahl, and Bilal}]{KeoghArXiV2025}
\bibinfo{author}{M.~Keogh}, \bibinfo{author}{J.~McTighe}, \bibinfo{author}{J.~Dahl}, \bibinfo{author}{O.~R. Bilal},
\newblock \bibinfo{title}{Experimental observation of flow instability control by metamaterial subsurfaces},
\newblock \bibinfo{journal}{arXiv preprint arXiv:2504.02053}  (\bibinfo{year}{2025}).
\bibitem[{Wiberg et~al.(2025)Wiberg, Park, Ramakrishnan, Saxton-Fox, Matlack, and Ansell}]{WibergAIAA2025}
\bibinfo{author}{D.~D. Wiberg}, \bibinfo{author}{S.~Park}, \bibinfo{author}{V.~Ramakrishnan}, \bibinfo{author}{T.~Saxton-Fox}, \bibinfo{author}{K.~Matlack}, \bibinfo{author}{P.~J. Ansell},
\newblock \bibinfo{title}{One-way coupling of surface vibration and k{\'a}rm{\'a}n vortex street instability},
\newblock in: \bibinfo{booktitle}{AIAA AVIATION FORUM AND ASCEND 2025}, \bibinfo{year}{2025}, p. \bibinfo{pages}{3252}.
\bibitem[{Navarro et~al.(2025)Navarro, Balderas, LaLonde, Velasquez-Gonzalez, Hoffman, Combs, and Restrepo}]{NavarroMatter2025}
\bibinfo{author}{J.~D. Navarro}, \bibinfo{author}{D.~Balderas}, \bibinfo{author}{E.~J. LaLonde}, \bibinfo{author}{J.~C. Velasquez-Gonzalez}, \bibinfo{author}{E.~N. Hoffman}, \bibinfo{author}{C.~S. Combs}, \bibinfo{author}{D.~Restrepo},
\newblock \bibinfo{title}{Stabilization of hypersonic shockwave/boundary-layer interactions with phononic metamaterials},
\newblock \bibinfo{journal}{Matter}  (\bibinfo{year}{2025}).
\bibitem[{Lin et~al.(2024)Lin, Goza, and Bae}]{LinAIAA2024}
\bibinfo{author}{C.-T. Lin}, \bibinfo{author}{A.~Goza}, \bibinfo{author}{H.~J. Bae},
\newblock \bibinfo{title}{Active control for turbulent drag reduction by periodic blowing and suction},
\newblock in: \bibinfo{booktitle}{AIAA AVIATION FORUM AND ASCEND 2024}, \bibinfo{year}{2024}, p. \bibinfo{pages}{3636}.
\bibitem[{Lin et~al.(2026)Lin, Ramakrishnan, Goza, Matlack, and Bae}]{LinArXiV2026}
\bibinfo{author}{C.-T. Lin}, \bibinfo{author}{V.~Ramakrishnan}, \bibinfo{author}{A.~Goza}, \bibinfo{author}{K.~H. Matlack}, \bibinfo{author}{H.~J. Bae},
\newblock \bibinfo{title}{Weakly coupled fluid-structure interaction between wall-bounded turbulent flows and defect-embedded phononic subsurfaces},
\newblock \bibinfo{journal}{arXiv preprint arXiv:2604.10430}  (\bibinfo{year}{2026}).
\bibitem[{Benjamin(1960)}]{benjamin1960effects}
\bibinfo{author}{T.~B. Benjamin},
\newblock \bibinfo{title}{Effects of a flexible boundary on hydrodynamic stability},
\newblock \bibinfo{journal}{J. Fluid Mech.} \bibinfo{volume}{9} (\bibinfo{year}{1960}) \bibinfo{pages}{513--532}.
\bibitem[{Landahl(1962)}]{landahl1962stability}
\bibinfo{author}{M.~T. Landahl},
\newblock \bibinfo{title}{On the stability of a laminar incompressible boundary layer over a flexible surface},
\newblock \bibinfo{journal}{J. Fluid Mech.} \bibinfo{volume}{13} (\bibinfo{year}{1962}) \bibinfo{pages}{609--632}.
\bibitem[{Ramakrishnan et~al.(2026)Ramakrishnan, {Machado Burgos}, Park, Matlack, and Goza}]{RamakrishnanJFS2026}
\bibinfo{author}{V.~Ramakrishnan}, \bibinfo{author}{A.~{Machado Burgos}}, \bibinfo{author}{S.~Park}, \bibinfo{author}{K.~H. Matlack}, \bibinfo{author}{A.~Goza},
\newblock \bibinfo{title}{A framework to systematically study the nonlinear fluid-structure interaction of phononic materials with aerodynamic flows},
\newblock \bibinfo{journal}{J. Fluids Struct.} \bibinfo{volume}{145} (\bibinfo{year}{2026}) \bibinfo{pages}{104569}.
\bibitem[{Riley et~al.(1988)Riley, Gad-el Hak, and Metcalfe}]{riley1988compliant}
\bibinfo{author}{J.~J. Riley}, \bibinfo{author}{M.~Gad-el Hak}, \bibinfo{author}{R.~W. Metcalfe},
\newblock \bibinfo{title}{Complaint coatings},
\newblock \bibinfo{journal}{Annual Review of Fluid Mechanics} \bibinfo{volume}{20} (\bibinfo{year}{1988}) \bibinfo{pages}{393--420}.
\bibitem[{Gad-el Hak(1996)}]{Gad-el-Hak1996compliant}
\bibinfo{author}{M.~Gad-el Hak},
\newblock \bibinfo{title}{Compliant coatings: A decade of progress},
\newblock \bibinfo{journal}{Appl. Mech. Rev.} \bibinfo{volume}{49} (\bibinfo{year}{1996}) \bibinfo{pages}{S147--S157}.
\bibitem[{Williamson and Govardhan(2004)}]{williamson2004vortex}
\bibinfo{author}{C.~H. Williamson}, \bibinfo{author}{R.~Govardhan},
\newblock \bibinfo{title}{Vortex-induced vibrations},
\newblock \bibinfo{journal}{Annu. Rev. Fluid Mech.} \bibinfo{volume}{36} (\bibinfo{year}{2004}) \bibinfo{pages}{413--455}.
\bibitem[{Sarpkaya(2004)}]{sarpkaya2004critical}
\bibinfo{author}{T.~Sarpkaya},
\newblock \bibinfo{title}{A critical review of the intrinsic nature of vortex-induced vibrations},
\newblock \bibinfo{journal}{J. Fluids Struct.} \bibinfo{volume}{19} (\bibinfo{year}{2004}) \bibinfo{pages}{389--447}.
\bibitem[{Shelley and Zhang(2011)}]{shelley2011flapping}
\bibinfo{author}{M.~J. Shelley}, \bibinfo{author}{J.~Zhang},
\newblock \bibinfo{title}{Flapping and bending bodies interacting with fluid flows},
\newblock \bibinfo{journal}{Annu. Rev. Fluid Mech.} \bibinfo{volume}{43} (\bibinfo{year}{2011}) \bibinfo{pages}{449--465}.
\bibitem[{Brotnow et~al.(2026)Brotnow, Barnes, Vincent, Ramakrishnan, Matlack, and Ansell}]{BrotnowSSRN2026}
\bibinfo{author}{M.~Brotnow}, \bibinfo{author}{C.~Barnes}, \bibinfo{author}{T.~Vincent}, \bibinfo{author}{V.~Ramakrishnan}, \bibinfo{author}{K.~H. Matlack}, \bibinfo{author}{P.~Ansell},
\newblock \bibinfo{title}{Compressible linear stability analysis for passive transition delay using phononic materials},
\newblock \bibinfo{journal}{Available at SSRN 7198919}  (\bibinfo{year}{2026}).
\bibitem[{Goza and Colonius(2017)}]{goza2017strongly}
\bibinfo{author}{A.~Goza}, \bibinfo{author}{T.~Colonius},
\newblock \bibinfo{title}{A strongly-coupled immersed-boundary formulation for thin elastic structures},
\newblock \bibinfo{journal}{J. Comp. Phys.} \bibinfo{volume}{336} (\bibinfo{year}{2017}) \bibinfo{pages}{401--411}.
\bibitem[{Deymier(2013)}]{DeymierSpringer2013}
\bibinfo{author}{P.~A. Deymier}, \bibinfo{title}{Acoustic metamaterials and phononic crystals}, volume \bibinfo{volume}{173}, \bibinfo{publisher}{Springer science \& Business media}, \bibinfo{year}{2013}.

\end{thebibliography}

\end{document}